\documentclass[prb,showpacs,twocolumn,superscriptaddress]{revtex4}

\usepackage{amsmath,graphicx,latexsym,bm}

\begin{document}

\title{Electrical phase diagram of bulk BiFeO$_3$}

\author{Massimiliano Stengel}
\affiliation{ICREA - Instituci\'o Catalana de Recerca i Estudis Avan\c{c}ats, 08
010 Barcelona, Spain}
\affiliation{Institut de Ci\`encia de Materials de Barcelona 
(ICMAB-CSIC), Campus UAB, 08193 Bellaterra, Spain}

\author{Jorge \'I\~niguez}
\affiliation{Materials Research and Technology Department, Luxembourg
  Institute of Science and Technology, 5 avenue des Hauts-Fourneaux,
  L-4362 Esch/Alzette, Luxembourg}
\affiliation{Institut de Ci\`encia de Materials de Barcelona 
(ICMAB-CSIC), Campus UAB, 08193 Bellaterra, Spain}

\date{\today}

\begin{abstract} 
We study the electrical behavior of multiferroic BiFeO$_3$ by means of
first-principles calculations.
We do so by constraining a specific component 
of the electric displacement field along a variety of structural
paths, and by monitoring the evolution of the relevant physical
properties of the crystal along the way.
We find a complex interplay of ferroelectric, antiferroelectric and
antiferrodistortive degrees of freedom that leads to an unusually rich
electrical phase diagram, which strongly departs from the paradigmatic
double-well model of simpler ferroelectric materials. In particular,
we show that many of the structural phases that were recently reported
in the literature, e.g. those characterized by a giant aspect ratio,
can be accessed via application of an external electric field starting
from the $R3c$ ground state.
Our results also reveal ways in which non-polar distortions (e.g., the
antiferrodistortive ones associated with rotations of the oxygen
octahedra in the perovskite lattice) can be controlled by means of
applied electric fields, as well as the basic features characterizing
the switching between the ferroelectric and
antiferroelectric phases of BiFeO$_{3}$.
We discuss the multi-mode couplings behind this wealth of effects, 
while highlighting the implications of our work as regards both 
theoretical and experimental literature on BiFeO$_{3}$.
\end{abstract}

\maketitle


\section{Introduction}

Our current understanding of ferroelectric materials is based on the
concept of ``soft mode'', a polar phonon that is unstable in the
centrosymmetric reference phase and whose condensation leads to a
spontaneous macroscopic polarization in the ferroelectric
phase.\cite{chapter-11}
As a function of the soft mode amplitude, the energy landscape has
typically a double-well shape, with a negative curvature at the origin
due to the aforementioned instability, and a quartic (or higher order)
positive term stabilizing the two ferroelectric minima.
Recent research, however, has demonstrated that in many materials such
a picture is way too simplistic: additional degrees of freedom
often compete (or, sometimes, cooperate) with the polar modes in a
nontrivial way, substantially complicating the energy landscape and
consequently the phase diagram of the material.
In most perovskites, these additional modes consist in strain degrees
of freedom (deformation of the unit cell) and antiferrodistortive
(AFD) tilts of the oxygen octahedra; magnetism is also very important,
especially in the context of magnetoelectric
multiferroic materials; finally there are some notable cases (and more
are appearing weekly as this research topic has been gaining
considerable momentum recently) where antiferroelectric (AFE) modes
play an important role as well.
All these lattice distortions, taken individually, are nonpolar in
nature, but their coupling (either mutual or to the polar modes) often
leads to unusual and potentially useful physical properties, of
interest to both information and energy technologies.
Understanding such couplings is crucial to devising new materials with
enhanced properties, and to gaining new fundamental insight into the rich
physics that these compounds display.

BiFeO$_3$ is an especially attractive material in this context, and it
has been intensely studied in the past few years as it remains one of
the very few known room-temperature multiferroics (i.e. compounds
where ferroelectricity and magnetism coexist in the same phase).
In its ground-state phase, BiFeO$_3$ adopts a distorted perovskite
structure with $R3c$ symmetry, where the spontaneous ferroelectric
polarization coexists with antiphase AFD tilts, and both are oriented
along the [111] pseudocubic direction.
Recent studies, both experimental\cite{Zeches-09,Guennou-11} and
theoretical,\cite{Dieguez-11,Hatt-10,Wojdel-10,Yurong-12} have
revealed that BiFeO$_3$ is a true polymorphic material: in addition to
the ground-state $R3c$ structure, it can adopt a rich variety of
low-energy metastable phases. Some of these phases, characterized by
peculiar physical properties (e.g. a monoclinic structure with a giant
$c/a$ ratio, where $c$ and $a$ are the out-of-plane and in-plane
lattice parameters), can be stabilized via application of an epitaxial
strain in a thin-film sample.
Interestingly, in many of these metastable states, distortion patterns
(e.g.  AFE, or in-phase AFD) that are not present in the ground state
appear, pointing to a complex behavior that largely escapes the
traditional models of ferroelectrics.
Thus BiFeO$_3$ provides us with a unique playground to explore the
interplay of competing order parameters, how they interact with
external perturbations (e.g. strain, electric fields, etc.) and how
these interactions may lead to enhanced functional properties.

From the point of view of first-principles theory, a significant
number of studies have focused on using strain as a means to exploring
the available configuration space,\cite{Hatt-10,Wojdel-10,Yurong-12}
in order to simulate epitaxial thin-film growth conditions.
Such studies have been very successful at revealing key physical
features, e.g. the isosymmetric transition from the ground-state
rhombohedral structure to the ``supertetragonal'' phase under strong
in-plane compression.\cite{Hatt-10}
However, relying on just strain as an external parameter provides a
necessarily limited perspective of the energy landscape, especially in
the present case where many order parameters coexist.
Another option that has also been quite successful is the so-called
``effective Hamiltonian'' approach,\cite{Zhong-94,Kornev-07} which
consists in analyzing the phonon spectrum of the high-symmetry
reference phase and the energetics of the dominant structural
instabilities, and using such information to build an approximate
low-energy model of the system under consideration.
This allows to study the behavior of the material in a larger variety
of conditions, for example at finite temperature or under an applied
electric field.\cite{Lisenkov-09}
Such an approach is most accurate when the amplitude of the distortion
is relatively small, and the number of higher-order terms that one
needs to incorporate in the Hamiltonian is limited. In other cases,
e.g. perovskites based on the lone-pair-active cations Pb or Bi,
anharmonicities are much stronger, and writing a reliable functional
with a small number of terms (and degrees of freedom) becomes more
problematic.\cite{Dieguez-11}

Recently, methodologies have appeared that allow one to go beyond many
of the limitations described above, by directly controlling the
\emph{electrical} variables (polarization and electric field) in a
simulation of a crystalline insulator.\cite{Souza-02,Stengel-07}
In particular, the possibility of performing a calculation at a fixed
value of the electric displacement field (the so-called
``constrained-$D$'' technique)\cite{fixedd} enables the study of the
\emph{direct} interaction between polar and other (AFD, AFE, elastic,
magnetic) degrees of freedom. This is of great help for identifying
the microscopic couplings that govern the stability of a given phase,
understanding the nature of the switching paths (and potential
barriers) between two local minima, and ultimately enrichening our
perspective on the mechanisms that are most relevant for the
functional properties of interest (piezoelectricity,
magnetoelectricity).

Here we perform a detailed study of bulk BiFeO$_3$ by means of the
constrained-$D$ technique.
In particular, by considering four different paths in electric
displacement space, we develop a comprehensive map where most of the
relevant low-energy phases can be readily located, clarifying their
mutual relationship. Furthermore, the evolution of all relevant
degrees of freedom is studied along each path, providing unique
insight into their coupling to the polar vector.
We show that the paradigmatic double-well potential of simple
ferroelectrics becomes, in the case of BiFeO$_3$, either a three- or a
four-well potential curve depending on the chosen path, with several
phase transitions (either first- or second-order) occurring along the
way.
Of particular note, we identify an important region of the phase
diagram where the physics is dominated by strong
\emph{antiferroelectric} displacements of the Bi atoms, whose
switching behavior under an applied field is fundamentally interesting
even beyond the specifics of BiFeO$_3$. (In fact, we could not find
earlier works where the field-induced switching of an
antiferroelectric phase is studied from first principles, except for 
a couple of recent investigations based on approximate 
potentials.\cite{Xu-15a,Xu-15b})
Interestingly, we could also identify some ranges of $D$ values where
BiFeO$_3$ displays \emph{complex} tilts (i.e. an in-phase and
anti-phase AFD mode coexisting along the same axis), which appears to
be a rare occurrence in the physics of simple perovskites.

\section{Methods}

Our calculations are performed within the local density approximation
to density functional theory, with an on-site Hubbard $U$
($U=3.3$~eV) applied to the $3d$ orbitals of Fe.
The core-valence interactions are dealt with in the framework of the
projector-augmented wave method,\cite{Bloechl-94} with a plane-wave
cutoff of 50~Ry.
We use a 20-atom monoclinic BiFeO$_3$ cell, which we construct as
follows. Let ${\bf a}_{1,2,3}$ be the real-space translation vectors
of a hypothetical 5-atom perovskite cell; then the corresponding
20-atom cell is defined by $\bar{\bf a}_3 = 2{\bf a}_3$, $\bar{\bf
  a}_1 = {\bf a}_1 - {\bf a}_2$, and $\bar{\bf a}_2 = {\bf a}_1 +
{\bf a}_2$.
(Note that, even if our calculations are always performed with the 20-atom 
cell, in some cases we find it convenient for presentation purposes 
to convert our data into the 5-atom cell representation by inverting 
the above formulas; such instances are clearly marked in the text.)
The larger $\sqrt{2} \times \sqrt{2} \times 2$ cell allows us to describe, 
in addition to the ferroelectric polarization (${\bf P}$), the magnetic order 
(we assume G-type antiferromagnetism throughout\footnote{The small energy gain
  that can be achieved by switching to a C-type spin arrangement in
  the phases with large aspect ratio is irrelevant in the context of
  the phenomena described here.}) and the relevant structural
distortions. These are in-phase antiferrodistortive tilts of the
oxygen octahedral network along $z$ (AFD$_{z}^+$), anti-phase AFD
modes along all Cartesian axes (AFD$_{x,y,z}^-$) and possible
antiferroelectric modes.
The Brillouin zone of the 20-atom cell is sampled with a special
$k$-point set that is equivalent to a $4 \times 4 \times 4$
Monkhorst-Pack sampling of the primitive (5-atom) perovskite unit.

The fundamental variable in the context of this work is the electric
displacement vector, ${\bf D}$,
$${\bf D} = \epsilon_0 \bm{\mathcal{E}} + {\bf P}.$$
In practice, in the BiFeO$_3$ results presented here, the contribution
of the electric field to ${\bf D}$ is always negligible compared to
that of ${\bf P}$; therefore, when looking at the graphs, we can
assume that we are essentially fixing the corresponding polarization
component to a given target value.

We perform our calculations by constraining only one component of the
reduced (see Section~\ref{reduced}) electric displacement field
$\hat{\bf D}$ at the time, thus mimicking a parallel-plate capacitor
with a fixed free charge at the surface. The choice of the component
of $\hat{\bf D}$ corresponds to fixing a certain crystallographic
orientation for the surface. In this work we shall work with either
the $\hat{D}_{001}$ (``out of plane'', also indicated as $\hat{D}_z$) 
or the $\hat{D}_{110}$ (``in plane'', also indicated as $\hat{D}_{xy}$)
component, corresponding to the $[001]$ and $[110]$ pseudocubic
directions, respectively.
The calculation is repeated several times while varying the relevant
component of $\hat{\bf D}$ (and hence while describing a path in
electric displacement space); at each point, the energy, internal
electric field, and all structural distortions of interest are
monitored and analyzed.
Note that we often find several coexisting phases at certain values of
$\hat{D}$, and sometimes the structure undergoes an abrupt
change within a small interval of $\hat{D}$; whenever such
changes are discontinuous, we shall identify them as
\emph{first-order} phase transitions, \emph{second-order} otherwise.

\subsection{Note on the parameters used in the analysis}
\label{reduced}

Since the system undergoes substantial relaxations of the cell shape
and volume as a function of ${\bf D}$, it is extremely convenient~\cite{fixedd} 
to represent all quantities (electrical and structural) in reduced
coordinates.
First, in addition to the real-space translation vectors, it is
helpful to introduce the ``dual'' reciprocal-space vectors, $\bar{\bf
  b}_{1,2,3}$, in such a way that
$$\bar{\bf a}_{i} \cdot \bar{\bf b}_{j} = \delta_{ij}.$$
Then, in close analogy with the definition of the reduced atomic
coordinates, we introduce the \emph{reduced} electric displacement
field and polarization (both with the dimension of a charge) as
$$\hat{D}_i = \Omega \bar{\bf b}_{i} \cdot {\bf D}, \qquad \hat{P}_i =
\Omega \bar{\bf b}_{i} \cdot {\bf P}.$$
The reduced electric field is, instead, written as
$$\hat{E}_i = \bar{\bf a}_{i} \cdot \bm{\mathcal{E}}$$
and has the dimension of a voltage.
In the remainder of this work we shall indicate the relevant components 
of $\hat{\bf D}$ by $[ijk]$ or $xyz$ subscripts, which better clarify 
their orientation with respect to the pseudocubic axes. One should keep
in mind, however, that the $\bar{\bf b}_{i}$ and $\bar{\bf a}_{i}$ 
triads that appear in the above definition are rotated, as they refer to the 
20-atom cell that we used in the calculations.
Whenever needed, one can easily convert between the two references
(such a conversion will be performed systematically, e.g. when presenting 
the results for the ``rigid-ion'' polarization, which we shall introduce
shortly).
It is important to stress, in this context, that the ``reduced'' electrical
variables that we just introduced are \emph{extensive} physical quantities, 
i.e. they depend both on the choice of unit cell axes and on cell size. 
The simulated cell can be simply thought as a parallel-plate capacitor: 
$\hat{E}$ is the applied voltage, and $\hat{D}$ is the total polarization
and displacement charge that has flowed through the cell facet. Both 
evidently grow with the size of the system, if the applied electric
field is kept constant.

In addition to the energy, internal field, electric displacement and
cell parameters, we further analyze our structures by extracting the
amplitude of four inequivalent AFD distortions (see figure captions)
and the reduced ``rigid-ion'' polarization.
The latter is a vector sum of the reduced atomic positions in the
20-atom cell, each weighted by the nominal ionic charge of each specie
(+3 for Bi and Fe, $-$2 for O). From this value we remove a fixed number
of ``quanta of polarization'' in order to ensure that the result
vanishes in centrosymmetric phases.
The primary reason for considering this quantity and not the Berry
phase polarization is that the latter is only available along the
field direction in our in-house first-principles code.
This is not a major issue: despite of the quantitative discrepancy,
the qualitative behavior (especially in relation to symmetries) of the
rigid-ion ${\bf P}$ is analogous to that of the full Berry-phase ${\bf
  P}$.
Nevertheless, it should be kept in mind that genuine electronic
effects (e.g.  anomalies in the Born dynamical charge tensor, or
response of the electron cloud to the macroscopic electric field) are
absent from the rigid-ion ${\bf P}$.

\section{Results}

\begin{figure}
\begin{center}
\includegraphics[width=3in]{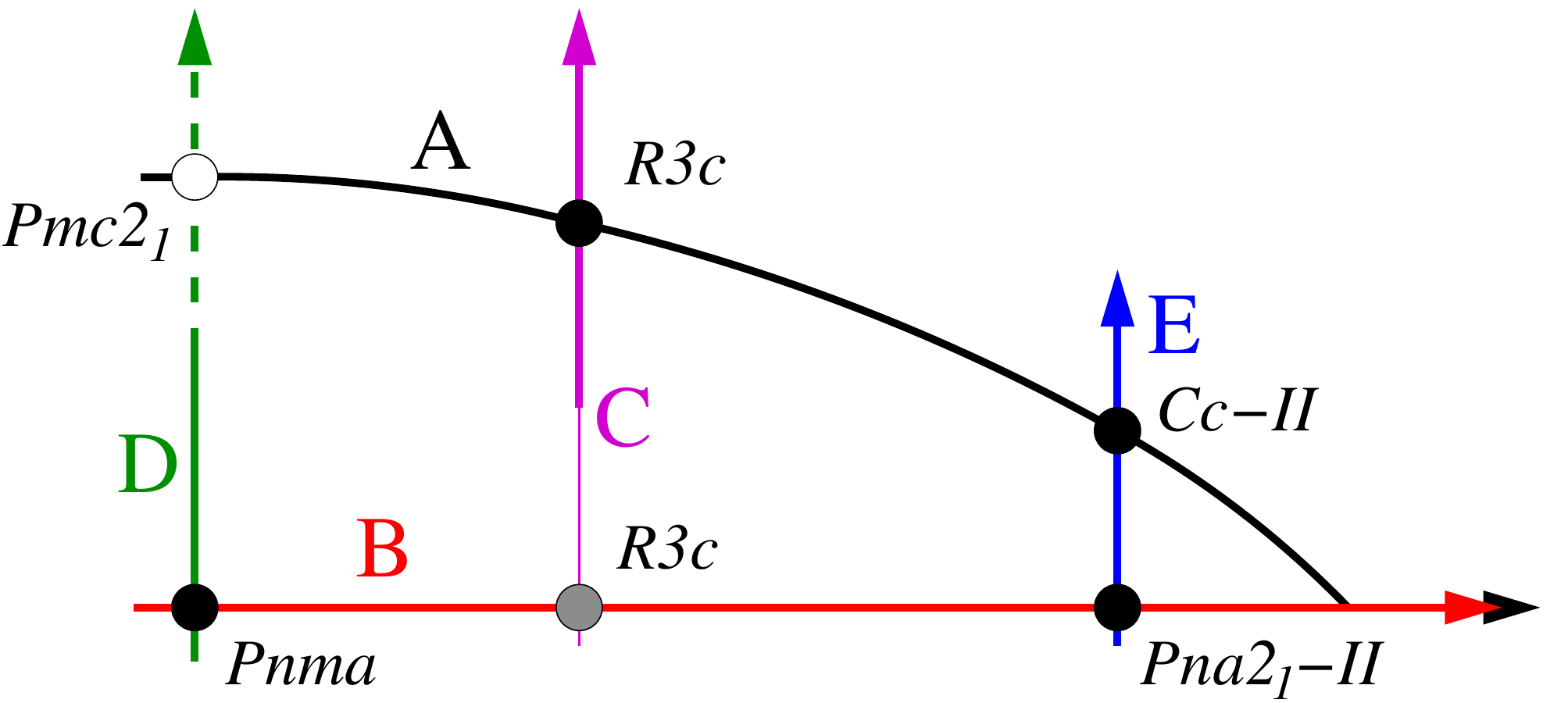} 
\end{center}

\caption{ \label{figpaths} Schematic phase diagram of BiFeO$_3$.  The
  horizontal and vertical axes correspond to $P_{001}$ and $P_{110} /
  \sqrt{2}$, respectively. The five different paths (A, B, C, D, E) in
  electric displacement space that were considered in this work are
  represented by lines/curves with arrows. The most relevant phases
  are indicated by circles (filled and empty circles are used,
  respectively, for local minima and saddle points).  The dashed
  portion of line D indicates that the corresponding path follows a
  potential ridge, rather than a valley (symmetry was imposed by hand
  to constrain the system along the $P_{001}=0$ line).  The thinner
  portion of line C indicates that the polarization acquires an
  out-of-plane component there; this path, therefore does \emph{not}
  cross path B -- the phase marked $R3c$ at $P_{110}=0$ bears a
  different color as a reminder.}
\end{figure}

\begin{table}
\begin{center}
\begin{ruledtabular}
\begin{tabular}{lcccc}
Phase        & $\Delta E$ &  $P_{xy}$ & $P_z$ & AFD pattern \\
\hline
%
$R3c$        & 0           &  76.8      &  54.2    & $(a^- a^- a^-)$ \\
$Cc$-II      & 98          &  42.1      & 140.1    & $(a^- a^- c^-)$ \\ 
$Pnma$       & 26          &  0         & 0        & $(a^- a^- c^+)$ \\
$Pna2_1$-II  & 101         &  0         & 141.5    & $(a^- a^- c^+)$ \\
$Pmc2_1$-II  & 106         & 116.1      & 0        & $(a^0 a^0 c^+)$ \\
$Pmc2_1$ (*) & 69          &  83.4      & 0        & $(a^- a^- c^+)$ \\
$Ima2$ (*)   & 72          &  94.0      & 0        & $(a^- a^- c^0)$ \\
\end{tabular}
\end{ruledtabular}
\end{center}
\caption{Relative energy, spontaneous polarization and AFD distortion
  pattern of the main phases of BiFeO$_3$ discussed in this work. The
  energy ($\Delta E$), in meV per formula unit, is referred to the
  $R3c$ ground state. The polarization vector is reported in units of
  $\mu$C/cm$^2$ ($xy$ refers to the in-plane component, $z$ to the
  out-of-plane one). The notation for the AFD pattern follows that of
  Glazer.\protect\cite{Glazer-75} Star symbols in the first column
  indicate phase that are not local minima but saddle
  points. \label{tab1}}
\end{table}

\begin{figure*}
\begin{center}
\begin{tabular}{r r}
\includegraphics[width=3.18in]{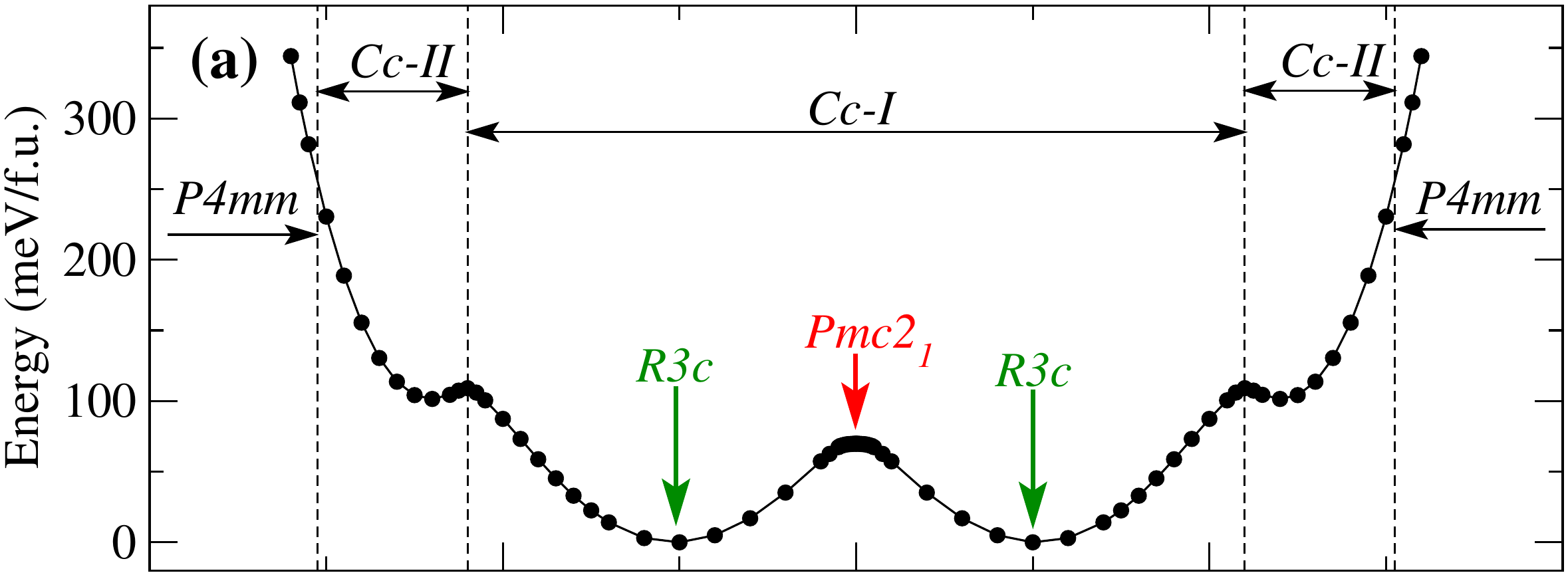}  & 
\includegraphics[width=3.1in]{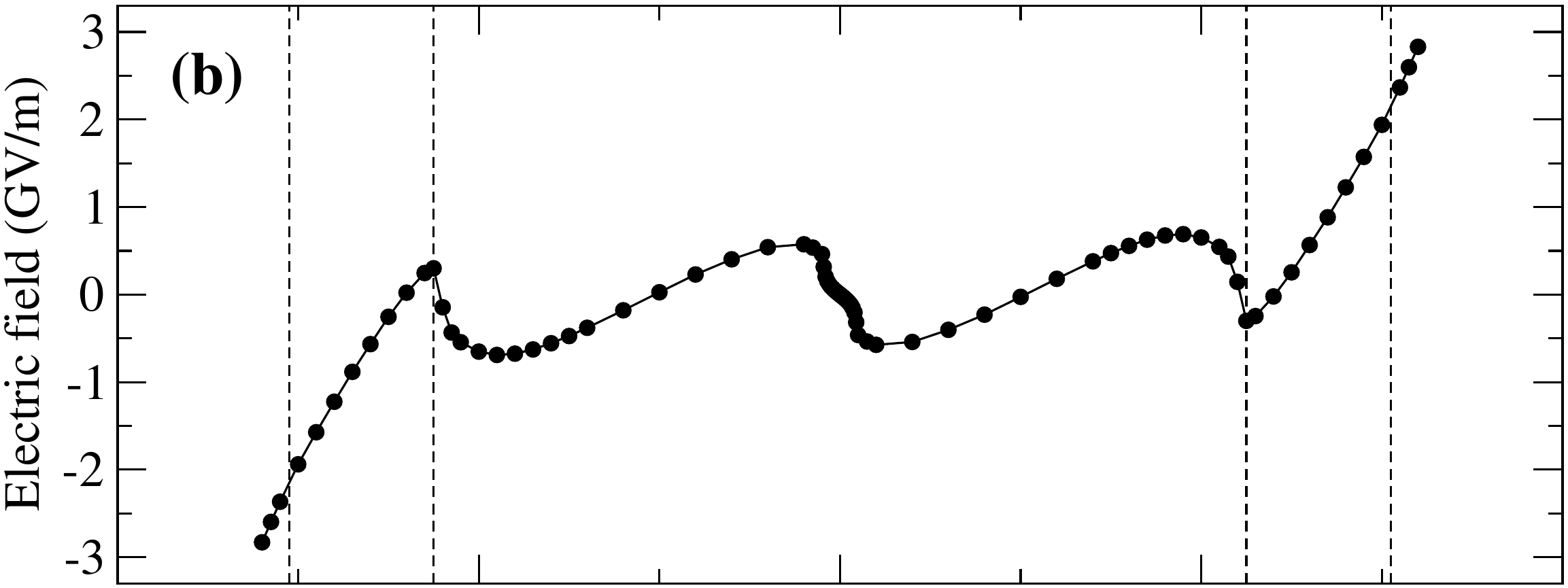} \\ 
\includegraphics[width=3.25in]{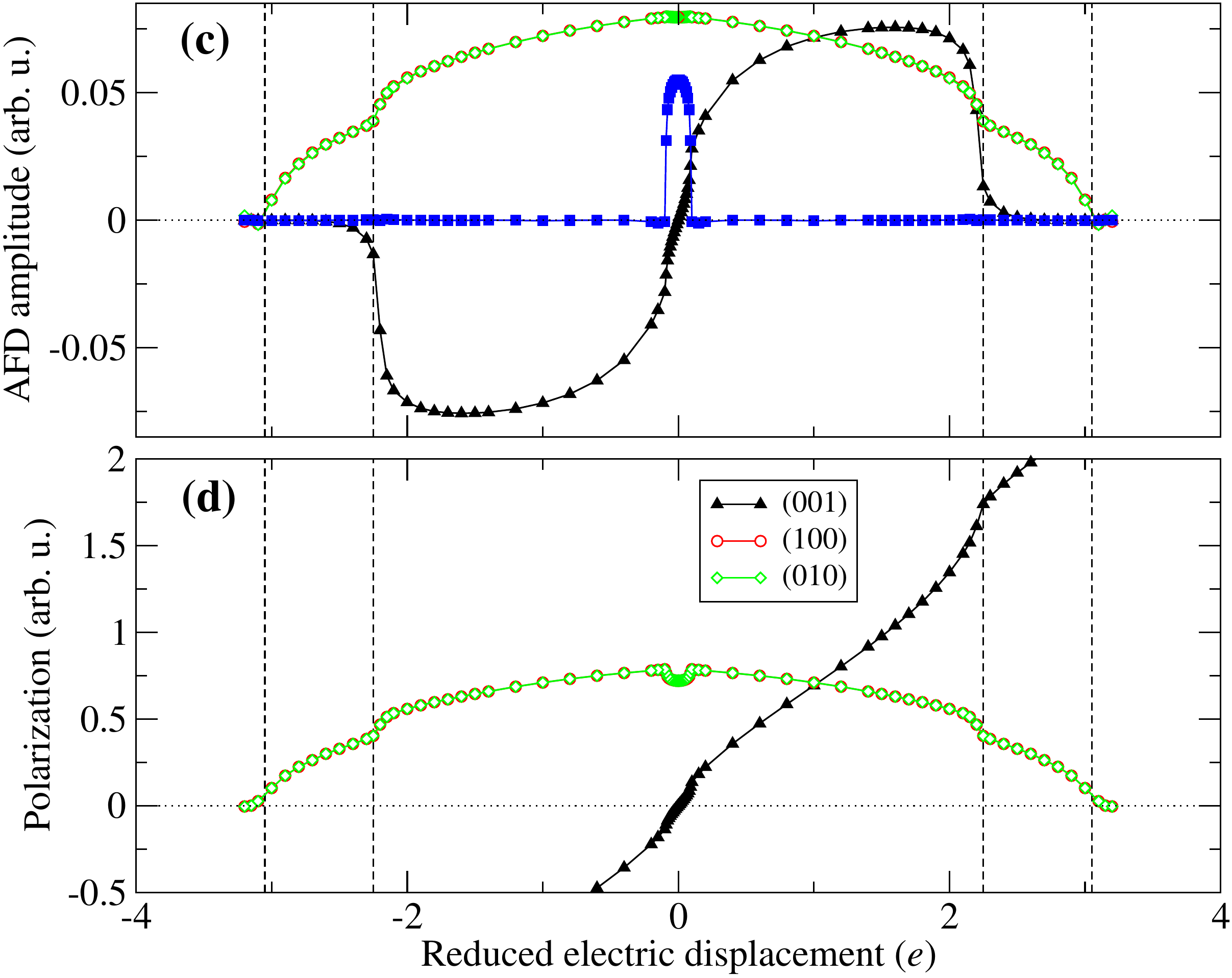} &
\includegraphics[width=3.18in]{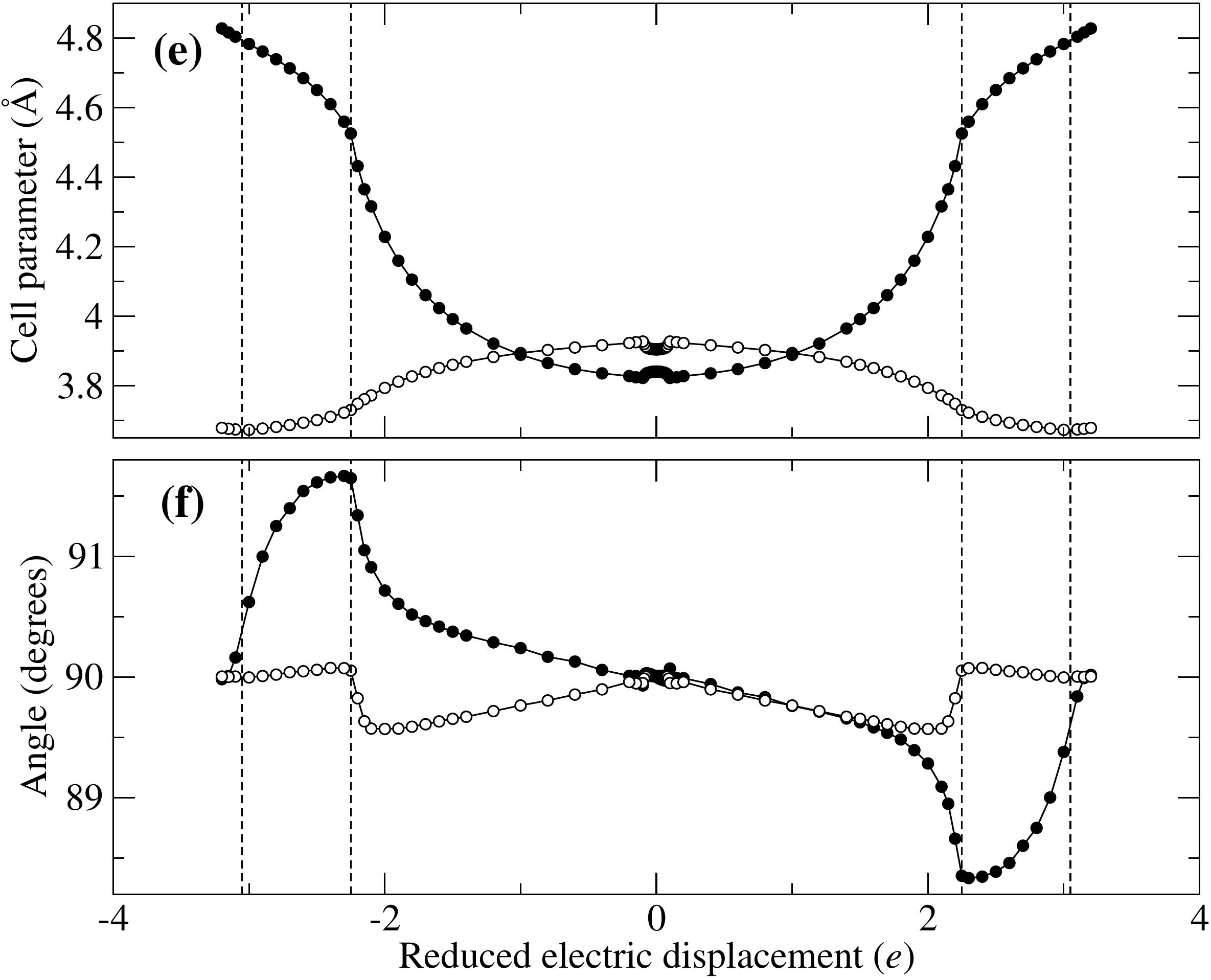} 
\end{tabular}
\end{center}

\caption{Evolution of the system along path A, obtained
by starting from the $R3c$ ground state of BiFeO$_3$ and varying 
$\hat{D}_{001}$.
(a): Energy (the zero is set to the the $R3c$ ground state);
(b): internal electric field; (c): anti-phase AFD$_{z}^-$ (filled black triangles), 
  AFD$_{x,y}^-$ (empty red circles and green diamonds, respectively), 
  and in-phase AFD$_{z}^+$ (filled blue squares) amplitudes,
(d): rigid-ion polarization vector (the same convention as
     in the AFD$^{-}$ case is used to label the individual $x$, $y$ and
     $z$ components);
(e): $|{\bf a}_3|$ (filled circles) and 
     $|{\bf a}_{1,2}|$ (empty circles); 
(f): angle formed by ${\bf a}_1$
  and ${\bf a}_2$ (empty circles), and by ${\bf a}_3$ and ${\bf a}_{1,2}$ 
                  (filled circles).
Data in (e-f) are referred to the 5-atom cell; values of the reduced electric 
displacement field, $\hat{D}_{001}$, are referred to the 20-atom cell. Vertical
dashed lines indicate second-order phase transitions.
  \label{fig:patha} }
\end{figure*}

In order to explore the phase diagram of BiFeO$_3$ as a function of
the electrical degrees of freedom, we investigate the evolution of the
equilibrium crystal structure along five selected paths in electric
displacement space, as schematically illustrated in Fig.~\ref{figpaths}.
These were obtained by using one of the lowest-energy (meta)stable
phases of BiFeO$_3$ as a starting point, and by subsequently
constraining either $\hat{D}_{001}$ or $\hat{D}_{110}$ to the
physically relevant range of values.
($\hat{D}_{001}$ or $\hat{D}_{110}$ indicate, respectively, the
  projection of the reduced electric displacement vector, $\hat{\bf
    D}$, along either the out-of-plane [001] or the in-plane [110]
  pseudocubic axis.)
The majority of the resulting phases (the main ones are explicitly 
indicated in Fig.~\ref{figpaths}) present a
$(\bar{1}10)$ mirror-symmetry plane; in such cases, $\hat{D}_{\bar{1}10}$
vanishes, and the resulting structures can be conveniently
mapped on a two-dimensional diagram spanned by $\hat{D}_{001}$ and
$\hat{D}_{110}$.
An exception to this rule occurs along path C, where the system acquires
a nonzero value of  $\hat{P}_{\bar{1}10}$, and hence breaks the
aforementioned mirror symmetry.

Before discussing the specifics of each path, it is useful to briefly
summarize (both for future reference and for comparison with earlier
works) the main phases of BFO that we have encountered in our study.
In Table~\ref{tab1} we report the energy (relative to the $R3c$ ground
state), spontaneous polarization and AFD distortion pattern
characterizing each phase.
The overall picture appears to be well in line with existing
literature data;\cite{Dieguez-11} there are some minor differences at
the quantitative level, which are most likely due to the choice of the
pseudopotential and/or other computational parameters.

\subsection{Path A}

\label{patha}










We start from the fully relaxed $R3c$ ground state of BiFeO$_3$ (at
$\hat{D} \sim 1.0$ in Fig.~\ref{fig:patha}) and vary $\hat{D}_{001}$,
i.e. apply an electric field along the out-of-plane pseudocubic
direction (corresponding to the reciprocal-space vector $\bar{\bf
  b}_3$).
At the energy minimum, the internal electric field [panel (b)] vanishes,
all AFD$^{-}$ [panel (c)] and ${\bf P}$ [panel (d)] components are equal,
as well as the cell parameters [(e)] and angles [(f)]; this is fully
consistent with the rhombohedral symmetry of the relaxed crystal.
In a vicinity of the global minimum the symmetry reduces to monoclinic
$Cc$ -- we shall refer to this region as $Cc$-I, in loose analogy to
earlier works. (In the recent literature, $Cc$-I has often been used
to indicate the phase of BiFeO$_3$ that results from the application
of an epitaxial strain; the present case of an electric field
perturbation at relaxed cell parameters yields a similar structure of
the same symmetry, hence our naming choice.) The symmetry breaking
produced by the field is obvious in all relevant degrees of freefom of
the system: The $\hat{P}_{001}$ component of the 
polarization grows at the expense of the in-plane one
[Fig.~\ref{fig:patha}(d)], and a similar behavior characterizes the
anti-phase AFD$^-$ vector [Fig.~\ref{fig:patha}(c)]; also, the $c/a$ ratio
increases, as well as the monoclinic angle
[panels (e) and (f), respectively].

At a sufficiently large field, BiFeO$_3$ undergoes an isosymmetric
transition to a structure with a much larger polarization and axial
ratio. The transition is accompanied by a drastic suppression of the
AFD$^{-}_{001}$ mode and a significant change in the monoclinic angle
of the cell. Such a structure has been the topic of several studies in
the recent past, and will be indicated as $Cc$-II henceforth. Note
that this structure is a local minimum, and that the transition is
continuous, i.e. of second-order character.
A further increase in the electric displacement field produces a
progressive increase of the aspect ratio and a reduction of the
AFD$^{-}_{110}$ amplitude and of the monoclinic angle.
At $\hat{D}_{001} \sim 3.1$, both AFD$^{-}_{110}$ and
$\hat{P}_{110}$ go to zero, and the crystal adopts the higher $P4mm$
(tetragonal) symmetry. Note that $P4mm$ is \emph{not} a metastable
state, nor even a saddle point in our simulations. In fact, such a
structure can be sustained only by an unrealistically large electric
field, at the limit of what one can afford even in a computer
simulation. We consider it unlikely that such extreme conditions may
be accessible in the laboratory. (Yet, it is possible to obtain such a
$P4mm$ phase by growing thin films on strongly compressive
substrates.\cite{Christen-11})

When moving towards smaller $\hat{D}_{001}$ values (which
corresponds to applying an $\mathcal{E}$-field antiparallel to
$\hat{P}_{001}$), the out-of-plane components of both the polarization
and the AFD vector are progressively suppressed, as expected, while
the in-plane components tend to a constant.
In a vicinity of the saddle point at $\hat{D}_{001}=0$, however,
something quite unexpected happens: The \emph{in-phase} AFD$_z^+$ mode
softens, eventually inducing a second-order phase transition to a
configuration where both AFD$_z^-$ and AFD$_z^+$ coexist. To explain
such an outcome, recall that AFD$_z^+$ is an active instability of the
cubic reference phase; yet, in the ground state $R3c$ phase such a
mode is stabilized by its strong biquadratic repulsion with
AFD$_z^-$. The strong suppression of AFD$_z^-$ at $\hat{D}_{001}
\sim 0$ brings AFD$_z^+$ back into play, explaining the second-order
transition. At this point, AFD$_z^-$ should become stable as the two
distortions are mutually exclusive; hovever, the existence of a small
residual out-of-plane component of ${\bf P}$ ``drives'' the AFD$^-$
vector to acquire an out-of-plane component as well, even if such a
distortion is no longer an active lattice instability. (The coupling
responsible for such an effect will be discussed in Section
\ref{couplings}.)
When $\hat{D}_{001}$ is forced to be exactly zero, the system
recovers a $(001)$ mirror plane, and the symmetry becomes
$Pmc2_1$. Note that a \emph{stable} phase of this same symmetry was
recently predicted to occur in BiFeO$_3$ thin films at very large
tensile strain. The present $Pmc2_1$ phase differs from the previously
reported one in that it is \emph{unstable} with respect to an
out-of-plane polar distortion, and has an aspect ratio that is close
to $1:1$. Thus, the present $Pmc2_1$ phase can be regarded, as far as
$\hat{P}_{001}$ is concerned, as a ``centrosymmetric'' reference
structure for the $R3c$ ground state. ($Pmc2_1$, of course, is not
centrosymmetric overall, as it is characterized by a large
$\hat{P}_{110}$.)

\subsection{Path B}

\label{pathb}

\begin{figure}
\begin{flushright}
\includegraphics[width=3.2in]{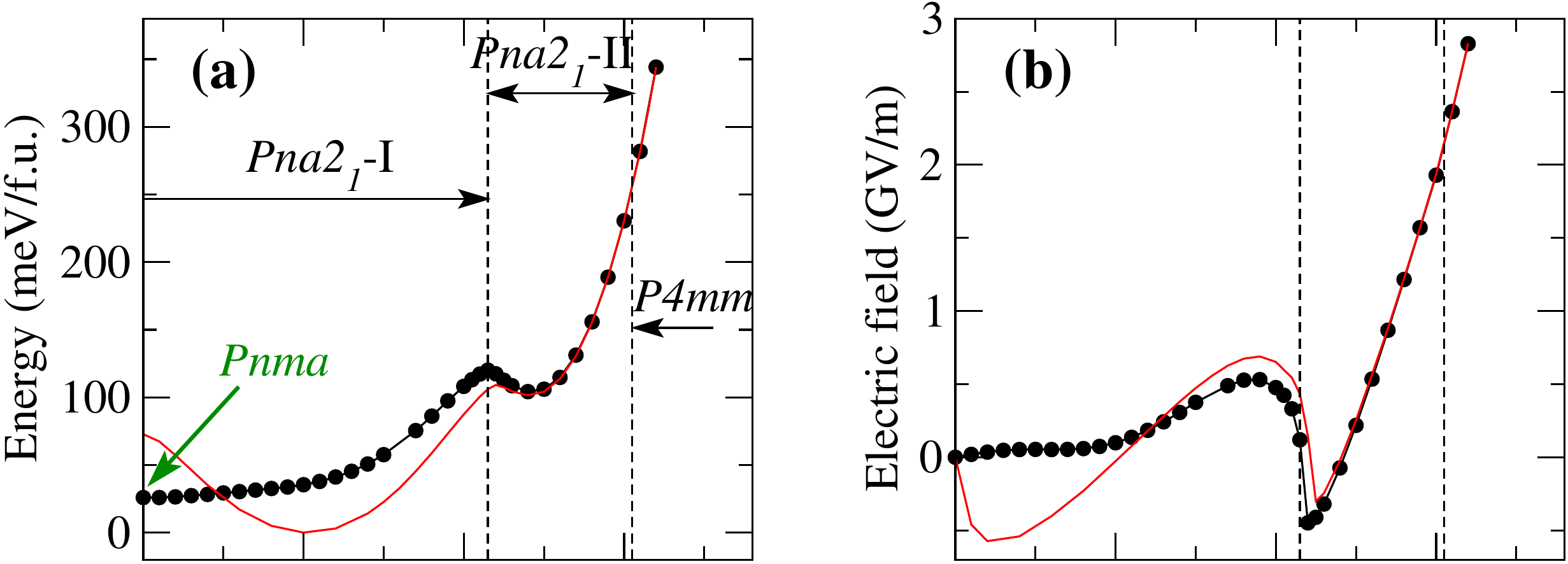} \\ 
\includegraphics[width=3.22in]{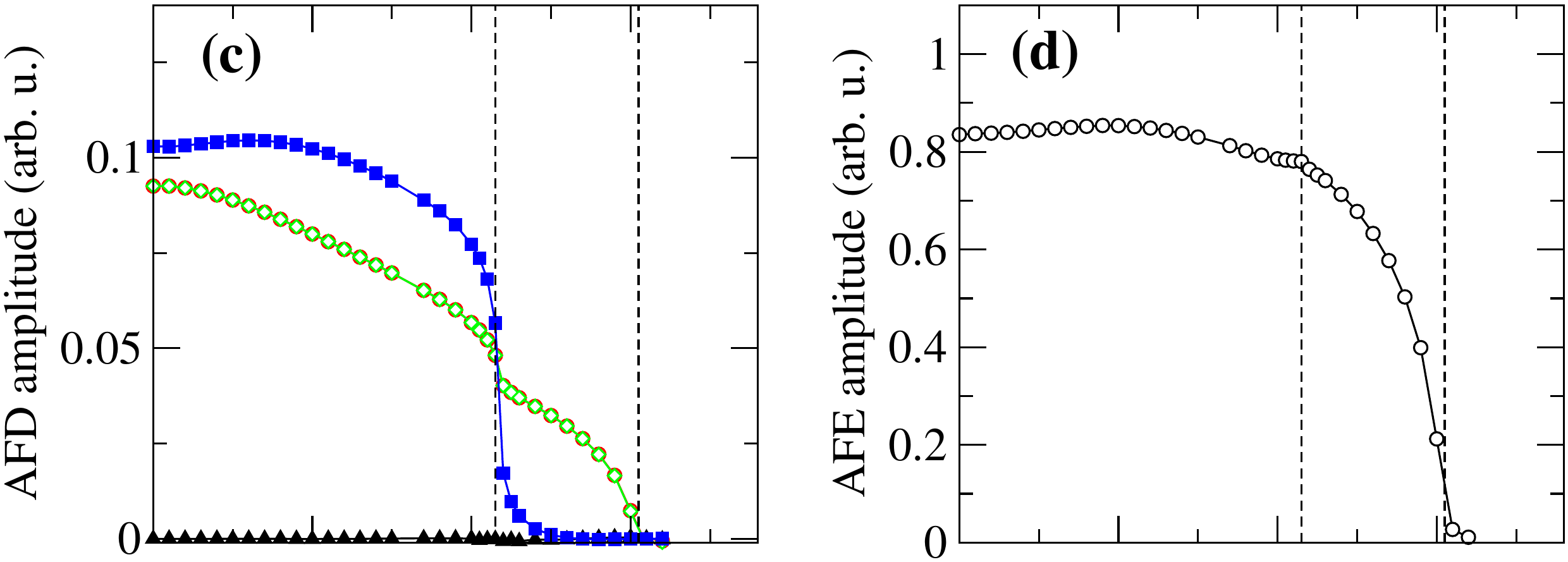} \\
\includegraphics[width=3.2in]{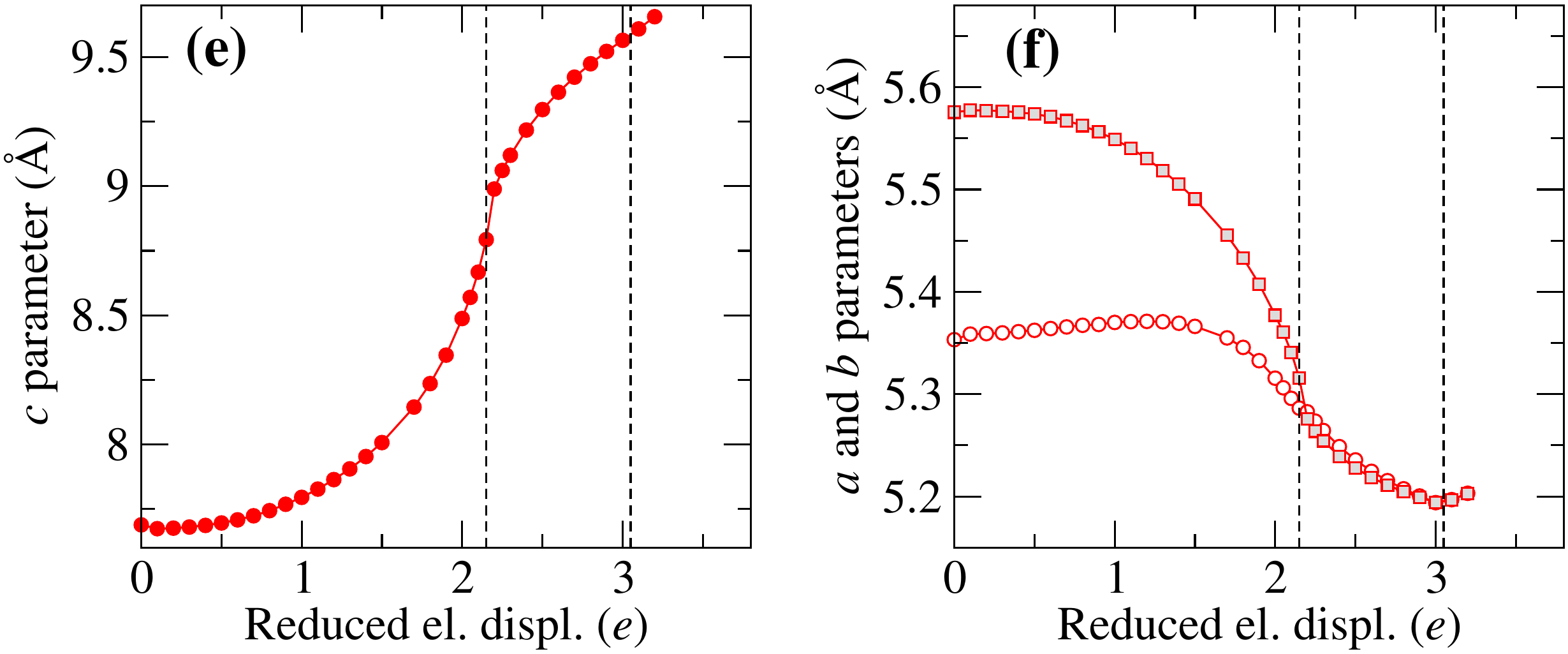}

\end{flushright}

\caption{Evolution of the system along path B, obtained
by starting from the $Pnma$ metastable state of BiFeO$_3$ (at $\hat{\bf D}=0$) and varying 
$\hat{D}_{001}$. 
(a): Energy (filled circles);
(b): internal electric field (filled circles); 
(c): anti-phase AFD$_{z}^-$ (filled black triangles), 
  AFD$_{x,y}^-$ (empty red circles and green diamonds, respectively), 
  and in-phase AFD$_{z}^+$ (filled blue squares) amplitudes,
(d): antiferroelectric (AFE) mode amplitude;
(e): $|\bar{\bf a}_3|$ (filled circles); 
(f): $|{\bf a}_{1}|$ (empty circles) and $|{\bf a}_{2}|$ (squares). 
Data in (e-f) are referred to the 20-atom cell (we use a red color to
distinguish them from the 5-atom cell data that are presented in other 
plots). Values of the reduced electric 
displacement field, $\hat{D}_{001}$, are referred to the 20-atom cell. 
Red curves without symbols in (a-b) refer to path~A [same
  as in Fig.~\ref{fig:patha}(a-b)], and are shown here for comparison.
Vertical dashed lines indicate second-order phase transitions. 
\label{fig:pathb}
}
\end{figure}

Excluding the $R3c$ structural ground state, the lowest-energy
(meta)stable configuration of BiFeO$_3$ is the $Pnma$
phase.\footnote{This is true when using the 
  local density approximation. Other energy functionals favor the 
  supertetragonal structures over the orthorhombic 
  one.\protect\cite{Dieguez-11}} 
It is natural, therefore, to perform an analogous computational
experiment as above (i.e. by controlling $\hat{D}_{001}$), but this
time using $Pnma$ as starting point. Note that the polarization
identically vanishes in the centrosymmetric $Pnma$ phase, so this
``path B'' in electric displacement space passes through the origin
(see Fig.~\ref{figpaths}). As we shall see, the system preserves two
mirror planes for all values of $\hat{D}_{001}$: The ${\bf P}$
vector stays parallel to the $z$ axis, and the lattice remains
orthorhombic.

In Fig.~\ref{fig:pathb}(a-b) we show the internal energy and electric 
field that we obtained along this path; for comparison, we also report 
the corresponding physical quantities that we have calculated along path~A
(see Section~\ref{patha}).
The electrical properties of the system look very different in the
central region of the phase diagram, i.e. for small to intermediate
values of $\hat{D}_{001}$.
At $\hat{D}_{001}=0$, the $Pnma$ state largely wins over the
unstable $Pmc2_1$ state discussed earlier. While increasing
$\hat{D}_{001}$, the energy difference between the two 
configurations becomes smaller, until the corresponding curves
sharply cross at $\hat{D}_{001} \sim 0.5$. For higher 
$\hat{D}_{001}$ values the path-A structures become favorable, 
until the curves merge at very large fields ($\hat{D}_{001} \sim 3.1$) 
into the same supertetragonal ($P4mm$) structure. 
Several things are interesting to note here, as we shall
illustrate in the following.

First, the energy as a function of $\hat{D}$ displays a remarkably flat 
plateau around $\hat{D}_{001} \sim 1.0$, pointing to an unusual dielectric
softness of the lattice in that region of parameter space.  
Closer inspection of the corresponding evolution of the electric field 
[Fig.~\ref{fig:pathb}(b)] shows a slight dip around the same value of 
$\hat{D}_{001}$ -- this indicates that the system is on the verge of 
having a metastable minimum there, of $Pna2_1$ symmetry and an out-of-plane 
${\bf P}$ of about 0.5 C/m$^2$.
Interestingly, earlier calculations indicate that a $Pna2_1$
metastable state with these characteristics indeed
exists.\cite{Dieguez-11} 
However, only gradient-corrected functionals seem to lead to a local 
minimum -- such a structure was not found whithin LDA, consistent with our 
present results.
%
%
This $Pna2_1$ state can be thought as the original $Pnma$ structure, 
with a AFD$_z^+$, AFD$_{xy}^-$ tilt pattern, plus a symmetry-lowering
out-of-plane polarization [see Fig.\ref{fig:pathb}(c)], which in 
our case is ``barely not spontaneous'' (i.e. it is induced by the electric field).

Second, note the presence of a strong \emph{antiferroelectric}
distortion [Fig.~\ref{fig:pathb}(d)] at essentially all values of
$\hat{D}_{001}$, except for the very largest ones ($\hat{D}_{001}
> 3.0$).
Such an AFE mode is, in many perovskites, a direct consequence of the
$Pnma$-like tilt pattern, which induces an antiparallel displacement
of the A-site cations (Bi in this case) in different (001) atomic
planes.\cite{Bellaiche-13}
The lone-pair activity of the Bi cations, however, leads to a
distortion amplitude that is unusually large (i.e. much larger that
one would expect if the AFE mode were only a secondary effect of the
tilts).
We shall go back to the specific role played by the AFE distortion in
the following Sections, where we study the effects of an in-plane
oriented electric field.

Finally, note the region $1.5 < \hat{D}_{001} < 3.0$. Here the
energetics and the electrical properties of the system
[Fig.~\ref{fig:pathb}(a-b)] qualitatively follow the same evolution regardless
of whether we are considering path~A or path~B.
%
%
%
Such an observation has two logical implications: (i) even if the
structures along paths A and B are quite different (in-plane ${\bf P}$
and AFD$_z^-$ in the former case, in-plane AFE and AFD$_z^+$ in the
latter), their response to an out-of-plane field (and therefore the
couplings between such distortions and $\hat{P}_{001}$) must be
comparable; (ii) the very similar energies of the two metastable
minima encountered along these paths, respectively $Cc$-II and
$Pna2_1$-II (the energy difference at the secondary minimum, 
$\hat{D}_{001}\sim 2.5$, approximately amounts to 3~meV per formula unit), 
suggest that it might be possible to switch between these
two phases upon application of an in-plane field; such a study will be
taken on in Section~\ref{pathe}.


\subsection{Path C}

\label{pathc}

\begin{figure}
\begin{flushright}
\includegraphics[width=3.2in]{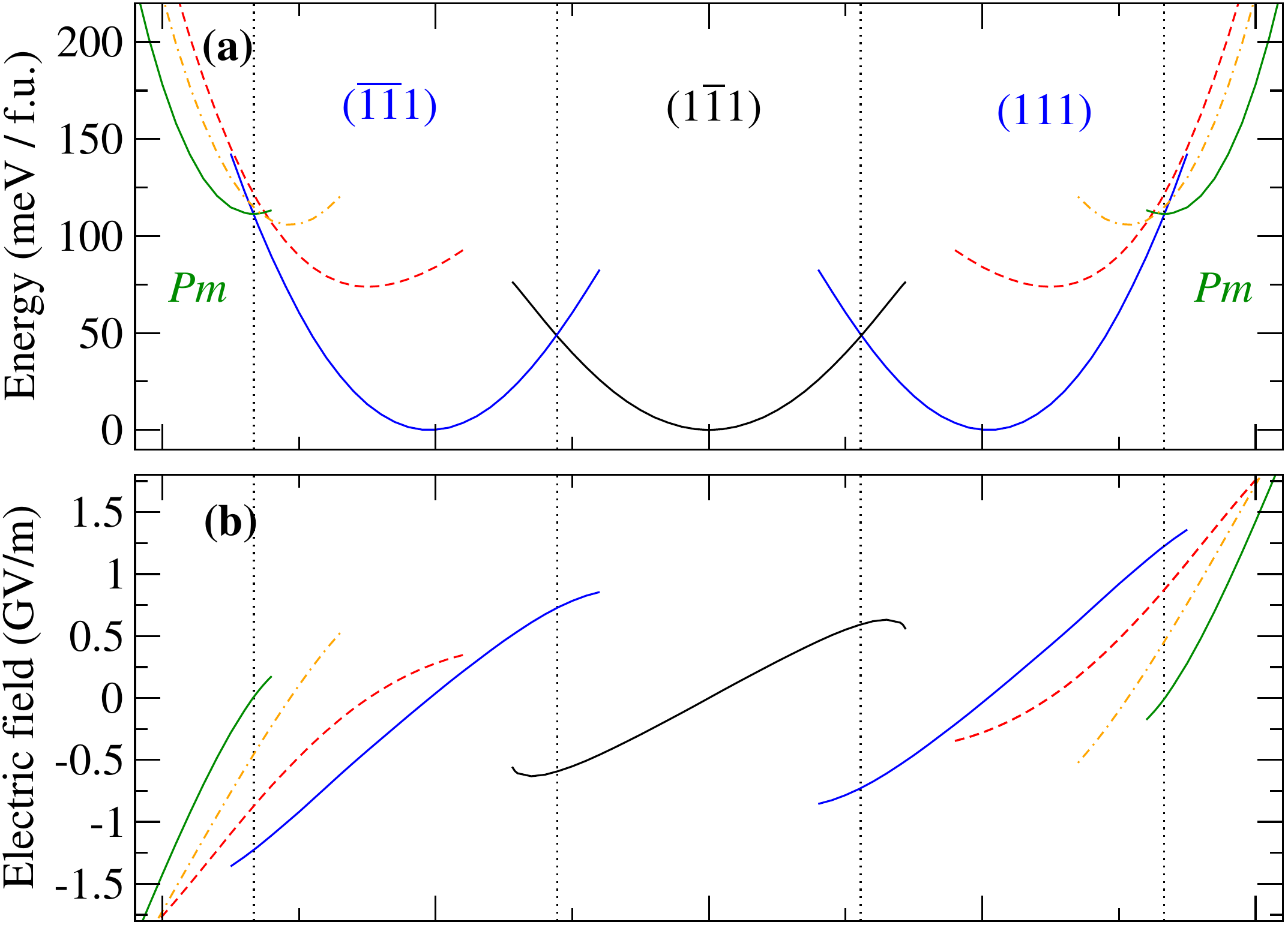} \\
\includegraphics[width=3.26in]{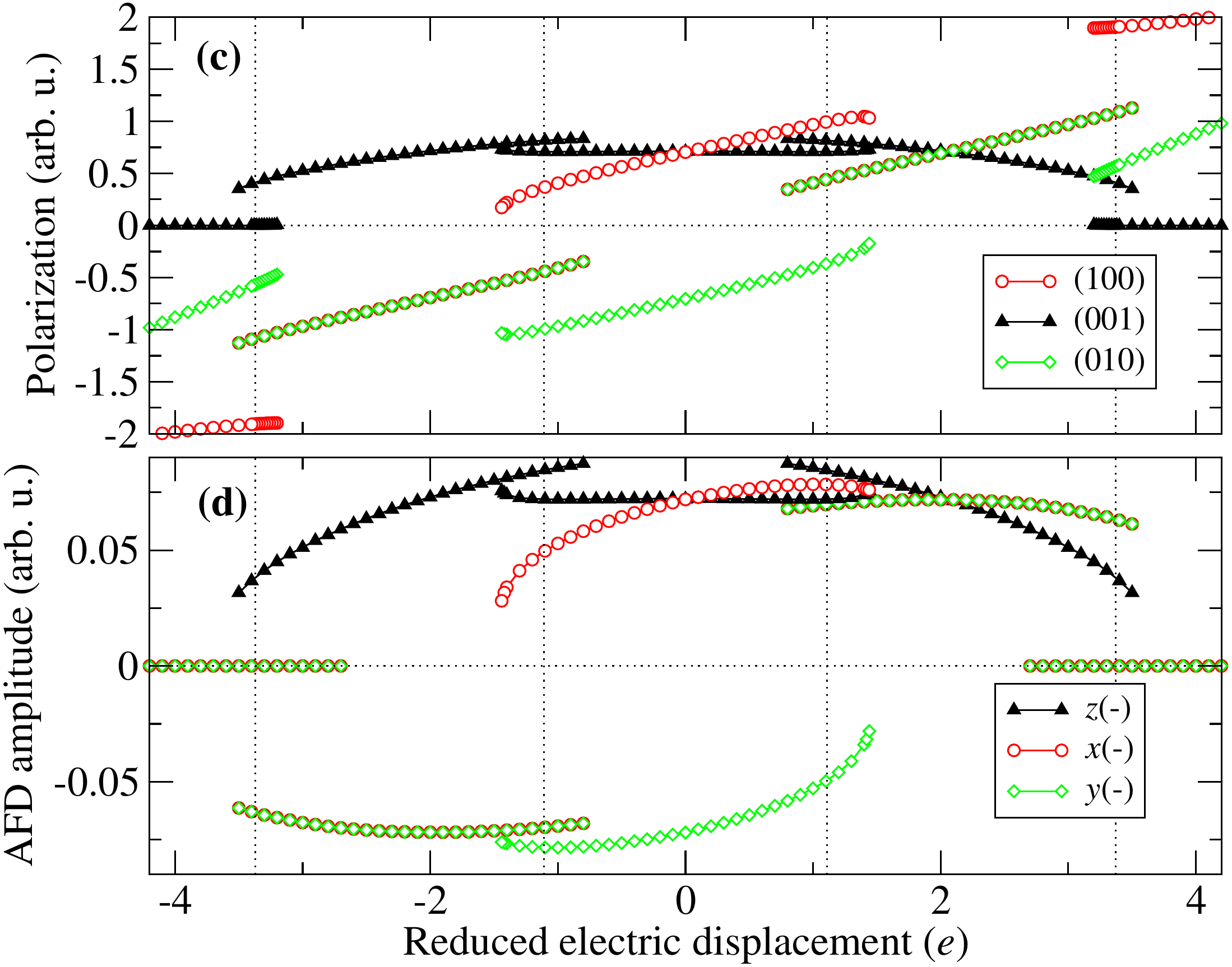}
\end{flushright}
\caption{
Evolution of the system along path C, obtained
by starting from the $R3c$ ground state, and varying 
$\hat{D}_{110}$ via an \emph{in-plane} electric field. 
(a): Energy (a different color or line style is used to 
     highlight different segments in configuration space);
(b): internal electric field; 
(c): anti-phase AFD$_{z}^-$ (filled black triangles), 
  AFD$_{x,y}^-$ (empty red circles and green diamonds, respectively);
(d): components of the rigid-ion polarization.
Vertical dotted lines indicate second-order phase transitions; 
horizontal dotted lines in (c) and (d) mark the zero of the graph. 
 \label{fig:pathc} The zero of the energy is set
  to the $R3c$ ground state, as usual. }
\end{figure}

%

In order to gain further insight on the electrical behavior of
BiFeO$_3$, it is particularly interesting to apply, starting from the
$R3c$ ground state, an electric field along the in-plane [110]
direction, rather than [001] as we did for path~A.
We expect this study to shed some light on how ferroelectric switching
proceeds in this material, e.g. whether the polarization vector
prefers to continuously rotate (as it is customarily assumed), or
rather jump at once from one orientation to the other.
It will be also interesting to clarify how the other order parameters
(i.e. those that are not directly acted upon by the electric field)
evolve during switching.
Of course, a realistic simulation of switching would require a much
more sophisticated theory, including temperature and disorder
effects. Here we focus on the (necessarily limited) information that
can be extracted from a hypothetical zero-temperature experiment,
where the sample is forced to remain single-domain throughout the
electrical cycle.

The internal energy and electric field as a function of the reduced
electric displacement, $\hat{D}_{110}$, are shown in
Fig.~\ref{fig:pathc}(a-b); the evolution of the internal degrees of freedom,
respectively polarization and AFD modes, is plotted in
Figs.~\ref{fig:pathc}(c) and \ref{fig:pathc}(d).
Note, in Fig.~\ref{fig:pathc}(a), the presence of three degenerate
global minima, all corresponding to the ground-state $R3c$ structure,
but with the polarization vector oriented in different directions: At
$\hat{D} \sim \pm 2$, ${\bf P}$ is parallel to [111], while at
$\hat{D} = 0$ it is oriented along [$1 \bar{1} 1$].
When the displacement field is varied around each of these minima, the
material behaves like a standard dielectric, with the internal
electric field growing linearly with $\hat{D}$ [black and blue curves
  in Fig.~\ref{fig:pathc}(b)].
The evolution of the internal degrees of freedom [Fig.~\ref{fig:pathc}(c-d)]
is, however, rather different depending on whether the electric field
acts \emph{perpendicular} to the spontaneous ${\bf P}$ (as in the
central region of the diagram) or at $\sim 35$ degrees to it (as in
the regions surrounding $\hat{D} \sim 2$).
At $\hat{D}=0$, we have a replica of the $R3c$ ground state, where
both ${\bf P}$ and the AFD vector are oriented along the $[1 \bar{1}
  1]$ direction; all components of the corresponding (pseudo)vector
quantities are equal in absolute value.
%
By applying an electric field along [110], $P_{001}$ 
remains roughly constant, while both $P_{100}$ 
and $P_{010}$ undergo a linear increase.
AFD$^-_z$ remains also constant, while the in-plane components
progressively rotate away from the $[1 \bar{1} 0]$ direction, moving
towards $[1 0 0]$; this behavior is in line with the known tendency of
the AFD vector to align with ${\bf P}$.
At large enough values of the displacement field (around $\hat{D} \sim
1.5$), the structure loses stability and an abrupt transition occurs
to the neighboring phase; note that the energy curves cross each other
somewhat earlier (at $\hat{D} \sim 1$), confirming the first-order
nature of the corresponding field-induced transition.
In the course of this transition, two main things happen: (i) the
$P_{1\bar{1}0}$ component [red curve in Fig.~\ref{fig:pathc}(c)] drops
to zero (i.e. $P_{100}$ and $P_{010}$ become equal); 
(ii) the $y$ component of the AFD$^-$ pseudovector [diamond
symbols in panel~(d)] switches sign, adopting the same value as
AFD$^-_x$. Both facts unambiguously indicate that the crystal has
switched to a different configuration. In fact, this structure reduces
to the [111]-oriented $R3c$ phase at $\hat{D} \sim 2.0$.
These results indicate that the polarization does not follow the
paradigmatic picture that is usually assumed for ferroelectric
materials, i.e. that of a smooth \emph{rotation} between one state and
the closest symmetry equivalent. The polar degrees of freedom seem to
abruptly jump, instead, between different configurations.
We ascribe this behavior to the peculiar chemistry of the Bi cation,
whose lone pair is prone to forming rather stiff bonds with the
neighboring atoms. It is reasonable to speculate that the Bi ions, as
they are pulled away from their equilibrium position by a strong
enough field, prefer to break free and directly switch site, rather
than gradually moving through the transition state. (Our conclusions
are in line with a recent first-principles investigation in which the
lowest-energy transition paths for ferroelectric switcihng in
BiFeO$_{3}$ were computed.\cite{Heron-14})

When forcing the in-plane polarization to larger values (i.e. at
either extreme, left or right, of the graphs in Fig.~\ref{fig:pathc}),
a new structure appears, of $Pm$ symmetry.
As one can see from Fig.~\ref{fig:pathc}(c-d), here all the AFD 
distortions disappear; conversely, there is a large in-plane polarization, 
with a larger component along $[100]$ and a smaller one along $[010]$.
While increasing $\hat{D}_{110}$ (and hence $\hat{E}_{110}$),
this phase again behaves like a linear dielectric material, as one can
appreciate from Fig.~\ref{fig:pathc}(b). 

The transition to the $Pm$ phase is another example of the peculiar
behavior of BiFeO$3$, which largely deviates from what one would
expect from common wisdom.
In fact, one would naively think that, by applying a [110]-oriented
field, one would end up increasing $P_{110}$ at the expense of
$P_{001}$ and, in parallel, increasing AFD$^{-}_{xy}$ at the expense of
AFD$^{-}_z$, until both order parameters eventually align with [110].
This would yield a hypothetical phase of $Ima2$ symmetry.
In order to check whether this scenario might apply to our case, we
constrained \emph{by hand} the symmetry of the system to $Ima2$ and
investigated how the resulting structure responds to a varying
electrical displacement field.
The results for the energy and internal electric field are shown as
dashed lines in Fig.~\ref{fig:pathc}(a-b).
Looking at the energy graphs, the high-$\hat{D}$ part of the blue
curve seems indeed to join smoothly the dashed red curve, as one would
expect from a second-order transition to $Ima2$.
However, before this happens, the system loses stability and
spontaneously falls into the $Pm$ state, which thereafter remains
stable at arbitrarily large values of the electric field.
As a result, the hypothetical transition to $Ima2$ never occurs,
contrary to the conclusions of Ref.~\onlinecite{Lisenkov-09}.

To further corroborate our finding of $Pm$ being the ground state at
large $\hat{D}_{110}$ values, we have examined one more alternative
structure, of $Pmc2_1$ symmetry. (We indicate this \emph{metastable}
phase as $Pmc2_1$-II, to distinguish it from the unstable $Pmc2_1$-I
phase that will be detailed in Sec.~\ref{pathd}.)
Our motivation for considering this trial structure comes from
Ref.~\onlinecite{Yurong-12}, where an analogous phase characterized by a
large [110]-oriented polarization and an AFD$^+_{z}$ distortion was
found to become favorable at large tensile strain.
As it can be appreciated from the corresponding curves in
Fig.~\ref{fig:pathc}(a) (orange dot-dashed), the energy is lower than that
of the $Ima2$ phase over the whole relevant range of
$\hat{D}_{110}$, confirming that $Pmc2_1$-II is indeed a valid
low-energy candidate.
However, the $Pm$ structure is clearly favored over $Pmc2_1$-II, in
agreement with the result of Ref.~\onlinecite{Lisenkov-09} that a $Pm$
structure becomes stable in the high-field regime.

\subsection{Path D}

\label{pathd}

\begin{figure}
\begin{center}
\includegraphics[width=3.2in]{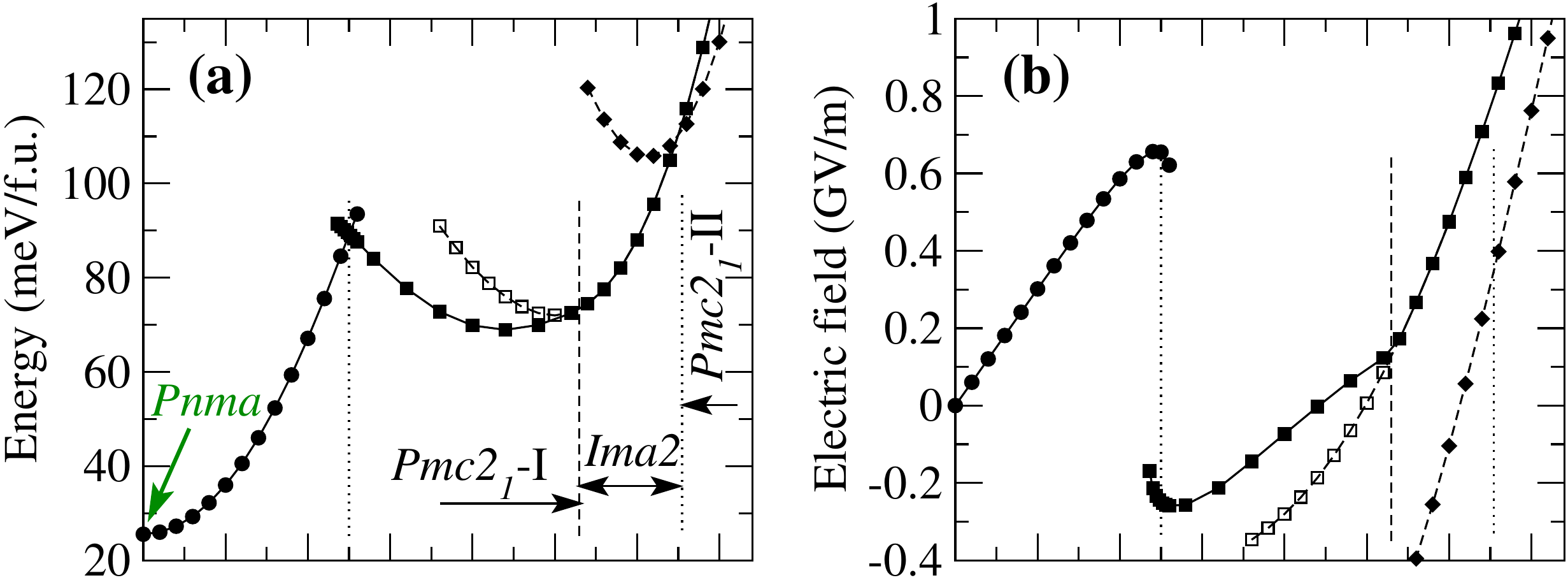} \\
\includegraphics[width=3.2in]{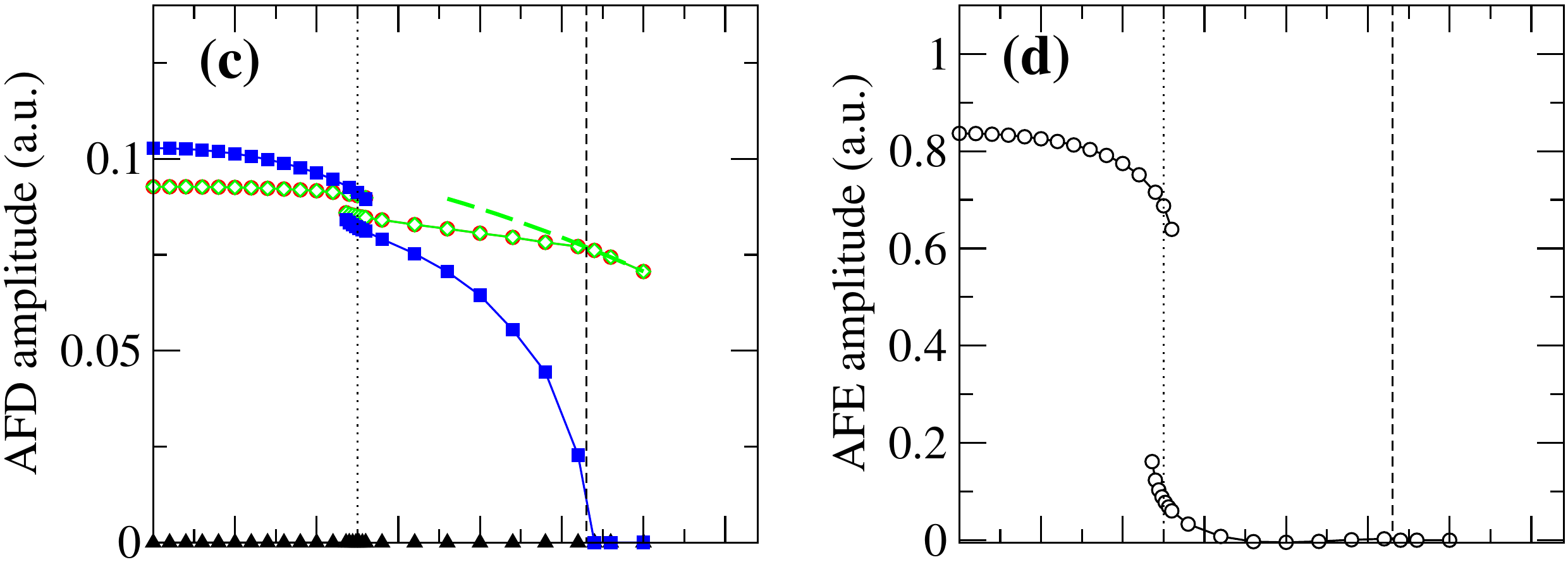} \\
\includegraphics[width=3.2in]{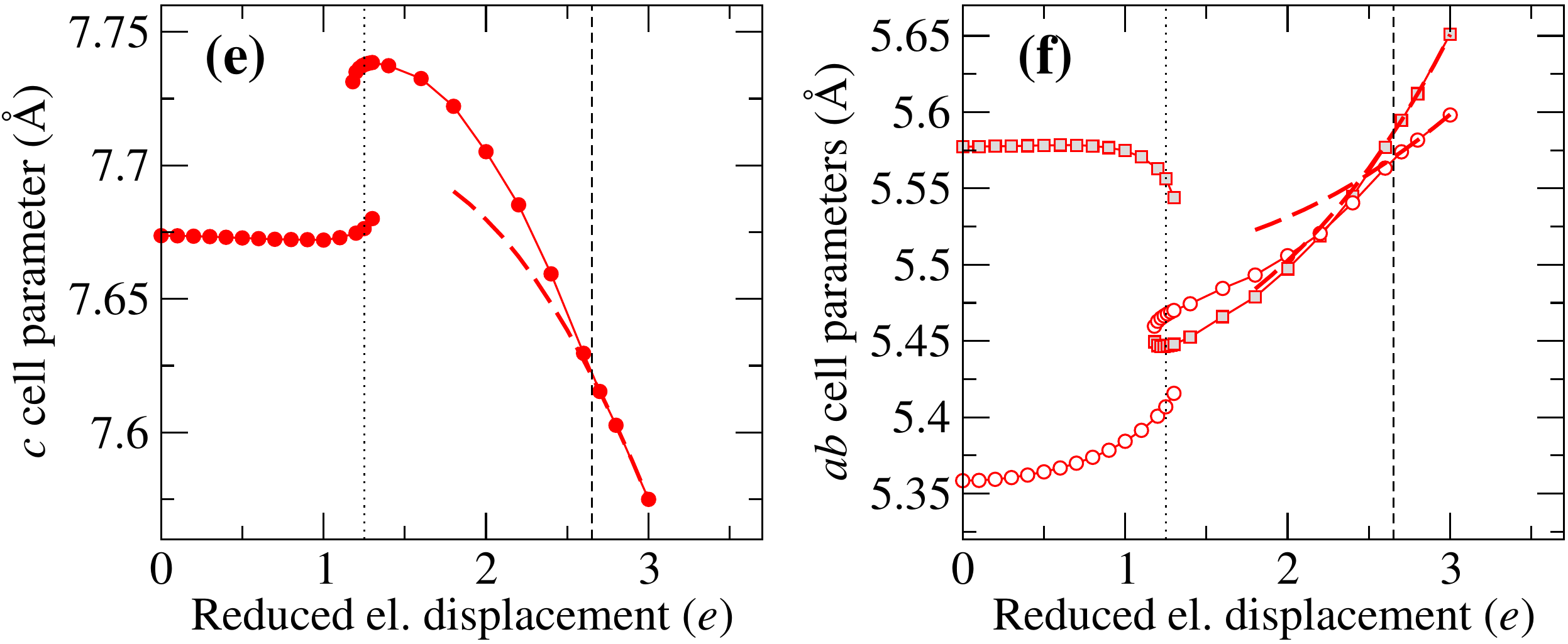}
\end{center}
\caption{
Evolution of the system along path D, obtained
by starting from the $Pnma$ metastable state of BiFeO$_3$ (at $\hat{\bf D}=0$) and varying 
$\hat{D}_{110}$ via an \emph{in-plane} electric field. 
(a): Energy. $Pmc2_1$ symmetry is imposed by hand in the data points shown as
  filled symbols; empty symbols refer to an analogous calculation
  where we impose the higher $Ima2$ symmetry; different symbols 
  (circles, squares, diamonds) are 
  used to highlight separate segments in configuration space.
(b): Internal electric field [same symbol convention as in panel (a)]. 
(c): Anti-phase AFD$_{z}^-$ (filled black triangles), 
  AFD$_{x,y}^-$ (empty red circles and green diamonds, respectively), 
  and in-phase AFD$_{z}^+$ (filled blue squares) amplitudes,
(d): Antiferroelectric (AFE) mode amplitude.
(e): $|\bar{\bf a}_3|$ (filled circles); the dashed line refers
to the calculations with imposed $Ima2$ symmetry.
(f): $|{\bf a}_{1}|$ (empty circles) and $|{\bf a}_{2}|$ (squares). 
Data in (e-f) are referred to the 20-atom cell.
Values of the reduced electric 
displacement field, $\hat{D}_{110}$, are referred to the 20-atom cell. 
Red curves without symbols in (a-b) refer to path~A [same
  as in Fig.~\ref{fig:patha}(a-b)], and are shown here for comparison.
Vertical dashed lines indicate second-order phase transitions. 
\label{fig:pathd}
  }
\end{figure}

The $Pnma$ phase described in Section~\ref{pathb} and the $Pmc2_1$
phase described in Section~\ref{patha} have several things in common.
In fact, all distortions are qualitatively similar, except that the
former has an AFE pattern in plane, while the latter is characterized
by an in-plane ${\bf P}$.
Based on this observation, it seems reasonable to suppose that one
could switch the AFE pattern of $Pnma$ into the FE pattern of $Pmc2_1$
by applying an in-plane field. What remains to be seen is whether this
will happen at physically realistic values of the electric field, 
$\mathcal{E}$, and if yes, how the AFD modes behave along this path.
(Note that this path does not always follow the bottom of a valley in
the two-dimensional electrical displacement diagram; the instability
of the $Pmc2_1$ state indicates that, at least in some segments of
this path, the system may follow a \emph{ridge}.
To prevent the system from falling into the adjacent basins, we impose
$Pmc2_1$ symmetry \emph{by hand} throughout the path.)

In Fig.~\ref{fig:pathd} we show the results
of the aforementioned computational experiment.
At small values of the in-plane field, the $Pnma$ phase behaves like a
linear dielectric, with the energy increasing quadratically as a
function of $\hat{D}_{110}$. [Note that the electric field, shown in
Fig.~\ref{fig:pathd}(b), is an almost perfectly linear function of
$\hat{D}_{110}$.]
In the same interval, the amplitude of AFD$_{z}^+$ undergoes a small
reduction, especially at larger fields, but remains large throughout;
the other AFD modes show little change.

At $\hat{D}_{110}$ values slightly larger than 1.0, an abrupt
transition occurs, where the electric field suddenly switches
sign. This transition is associated with the reorientation of the
electrical dipoles of the cell, whose ordering goes from antiferro to
ferro [the evolution of the AFE amplitude is shown in
Fig.~\ref{fig:pathd}(d)].
Remarkably, the AFD$_{z}^+$ mode survives even after the transition,
although at a slightly reduced amplitude. The antiferroelectric Bi
displacement pattern, therefore, is not necessarily a prerequisite to
having a AFD$_{z}^+$ distortion, although these two modes clearly
cooperate. Indeed, even after the transition to a ferroelectric
ordering, a small AFE component survives, most likely as a secondary
effect of the AFD$_{z}^+$ mode. (This is complementary to the example
of Section~\ref{pathe}, where a small AFD$_{z}^+$ amplitude develops
as a secondary effect of the large AFE distortion.)

By increasing the in-plane polarization the AFD$_{z}^+$ mode becomes
progressively weaker, until it disappears completely at $\hat{D}_{110}
> 2.7$. Here, the system undergoes a second-order transition to a
phase of (higher) $Ima2$ symmetry. (As we mentioned in the previous
section, this phase is, again, unstable -- it could occur in this
specific example solely because of the imposed symmetry constraints.)
If we perform a backward sweep starting from the $\hat{D}_{110} > 2.7$
region, and progressively decrease the field while imposing the $Ima2$
symmetry by hand, we obtain the energy and field values that are
represented as empty symbols in Fig.~\ref{fig:pathd}(a-b).
The difference between these new points and the original ones (which
were obtained within the lower $Pmc2_1$ symmetry) illustrates how much
energy is gained (and how the relevant properties of the system are
altered) by the condensation of the AFD$_z^+$ tilts.
[The evolution of the AFD$_{x,y}^-$ amplitude and of the cell
parameters in the $Ima2$ phase is shown as dashed curves in
Figs.~\ref{fig:pathd}(c) and~\ref{fig:pathd}(e-f).]

At even larger values of $\hat{D}$, the system eventually adopts the
$Pmc2_1$-II structure that we described earlier. (We omitted the
structural information about this phase in the figures, to avoid
overcharging them with redundant data.)
This phase is stable at large fields; its energy and internal field
are indicated by a dashed line with diamond symbols in
Fig.~\ref{fig:pathd}(a-b).

\subsection{Path E}

\label{pathe}

\begin{figure}
\begin{flushright}
\includegraphics[width=3.2in]{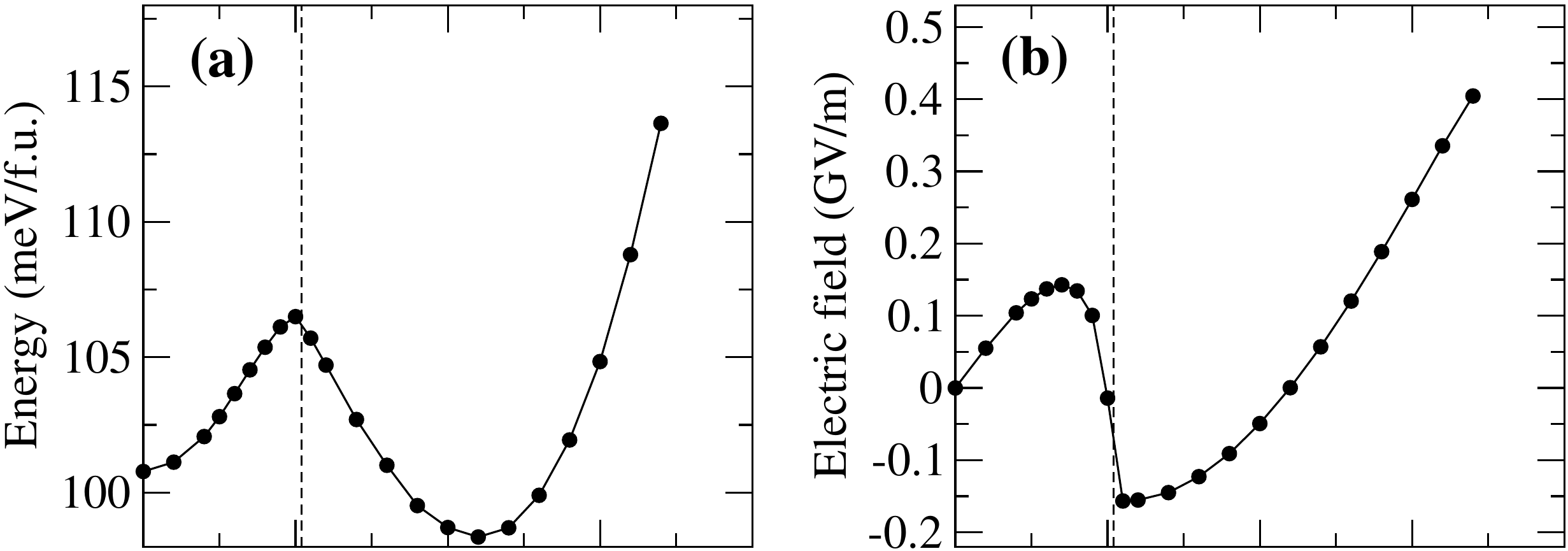} \\
\includegraphics[width=3.25in]{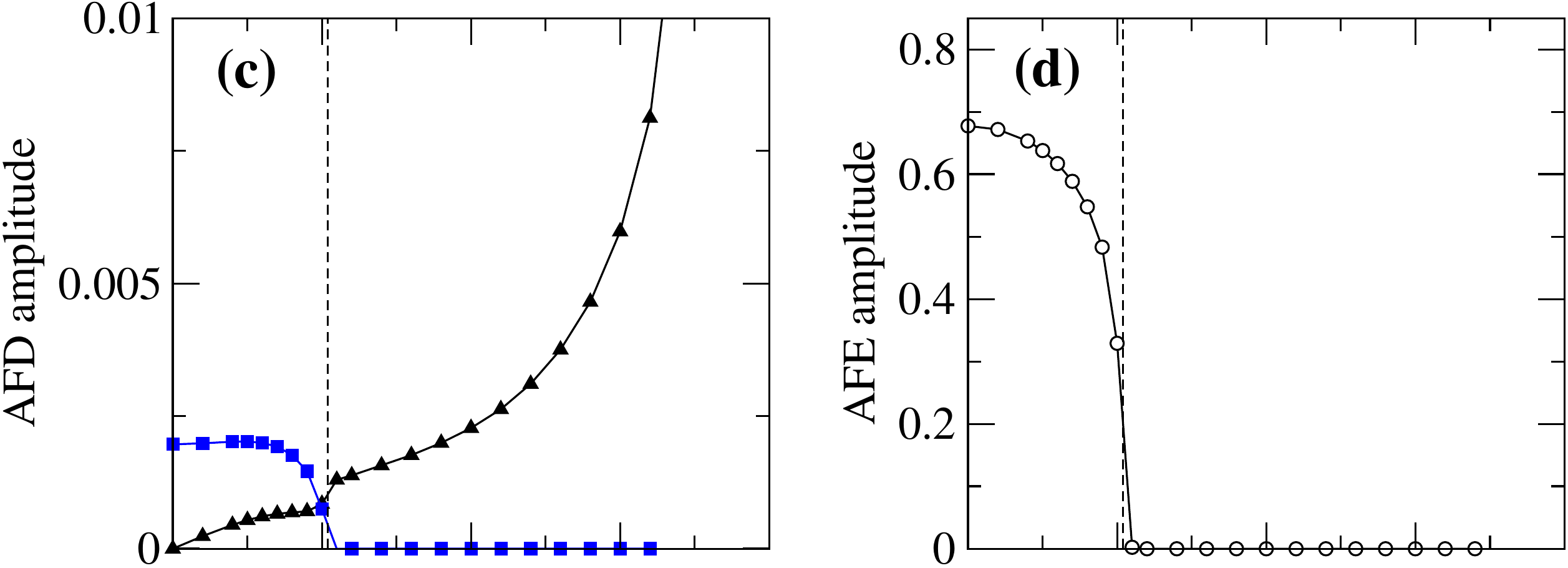} \\
\vspace{2pt}
\includegraphics[width=3.25in]{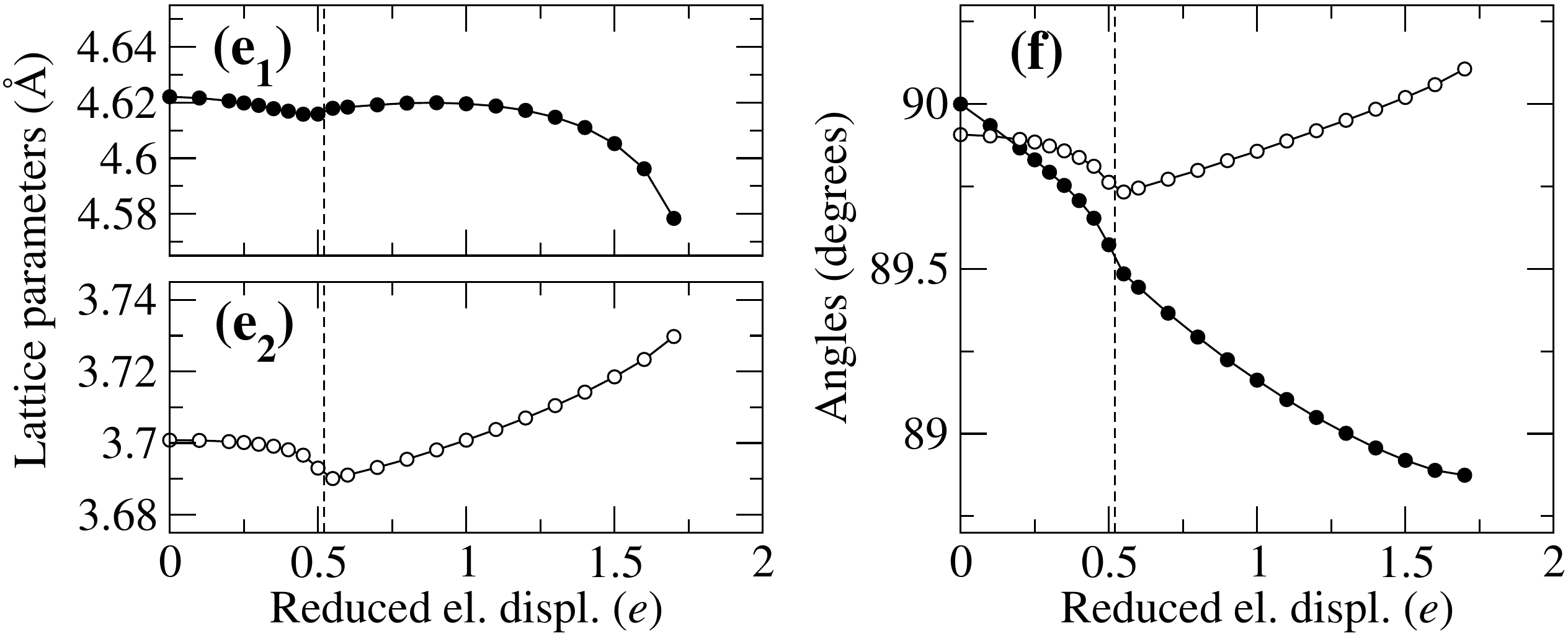}
\end{flushright}
\caption{
Evolution of the system along path E, obtained
by starting from the $Pna2_1$-II state, and varying 
$\hat{D}_{110}$ via an \emph{in-plane} electric field. 
(a): Energy.
(b): Internal electric field. 
(c): Anti-phase AFD$_{z}^-$ (filled black triangles), 
  and in-phase AFD$_{z}^+$ (filled blue squares) amplitudes 
  (AFD$_{x,y}^-$ mode amplitudes remain roughly constant throughout, 
   and are not shown).
(d): Antiferroelectric (AFE) mode amplitude.
(e): $|{\bf a}_3|$ (filled circles) and $|{\bf a}_{1,2}|$ (empty circles).
(f): angle formed by ${\bf a}_1$
  and ${\bf a}_2$ (empty circles), and by ${\bf a}_3$ and ${\bf a}_{1,2}$ 
                  (filled circles).
Data in (e-f) are referred to the 5-atom cell, as in Fig.~\ref{fig:patha}.
Values of the reduced electric 
displacement field, $\hat{D}_{110}$, are referred to the 20-atom cell. 
The vertical dashed line indicates a second-order phase transition. 
\label{fig:pathe}
}
\end{figure}

In Fig.~\ref{fig:pathb}(a) we have, at $ \hat{D}_{001} \sim $2.5, two almost
overlapping minima.
The lower one, of $Cc$ symmetry, is reached by progressively
increasing $\hat{D}_{001}$ from the ground-state $R3c$ phase (red
curve). It is characterized by a high aspect ratio with a noticeable
tilt of the $c$ axis [see Fig.~\ref{fig:patha}(f)], both in-plane and
out-of-plane components of the polarization [the latter is much
larger, see Fig.~\ref{fig:patha}(d)], and anti-phase AFD distortions along all
three directions (AFD$_{x,y}^-$ are both significant, while AFD$_z^-$
is small).
The other (3 meV higher in energy) local minimum, of $Pna2_1$ symmetry, is
reached when starting from $Pnma$ (black curve). It is similar to the
$Cc$-type minimum that we have just discussed, except for a few
important details: the $c$ axis is not tilted; the in-plane components
of ${\bf P}$ vanish identically, and are replaced by a significant
in-plane AFE distortion; the small (anti-phase) AFD$_{z}^-$ is
replaced by an equally small (in-phase) AFD$_{z}^+$.
If it weren't for such small AFD$_{z}$ components and for the AFE
pattern, one could think of the latter phase as the
``centrosymmetric'' reference structure for the former when the
polarization is switched in plane. If this were true, it would be
reasonable to expect the $Pna2_1$ phase to be a saddle point for
switching, and therefore be unstable.
This is, however, not the case: both phases are local minima. A
possible hypothesis to explain this outcome is that the either the
small AFD$_{z}^+$ distortion or the AFE mode play a key role in
stabilizing $Pna2_1$.

To shed some light on this point, and investigate whether (and how) it
may be possible to electrically switch the system between the
$Pna2_1$-II and $Cc$-II states, we have performed a study similar to
that of Section~\ref{pathd}, i.e. applying an in-plane field along
[110], but this time starting from $Pna2_1$-II.
As shown in Fig.~\ref{fig:pathe}(a), the energy first grows quadratically as
a function of $\hat{D}$ [note the initially linear evolution of
$\mathcal{E}$ in panel (b)], and a small AFD$_z^-$ component appears (as
a consequence of the tilting of the ${\bf P}$ vector away from the
vertical axis); interestingly, here (in close analogy to the situation
discussed in the context of path A) both AFD$_z^-$ and AFD$_z^+$
coexist for a certain range of $\hat{D}$ values.
At larger values of $\hat{D}_{110}$, the dielectric response of the
system progressively softens, and the AFE modes undergo an
increasingly drastic suppression.
Eventually, a second-order transition occurs that is driven by the
disappearence of the AFE modes. At the same time, the AFD$_z^+$ modes
also disappear (recall that they are a secondary effect of the AFE
distortion in this region of the electrical phase diagram). At even
larger fields, the AFD$_z^-$ modes progressively grow in magnitude,
and the system overall recovers a linear dielectric behavior.

Thus, we can conclude that the strong AFE distortions (the amplitude
of the AFE mode in $Pna2_1$ is comparable to that in the $Pnma$ phase:
Bi atoms move by about 0.19~\AA\ and 0.29~\AA\ from their ``ideal''
position in the $Pna2_1$ and $Pnma$ structures, respectively), rather
than the small AFD$_z^+$ modes drive the transition between
$Pna2_1$-II and $Cc$-II.
One can think in the following terms. Imagine a hypothetical
``reference'' structure, obtained from the $Pna2_1$ by removing both
the AFE and AFD$_{z}^+$ modes.
Such a structure has an unstable polar \emph{branch}, reflecting the
marked tendency of the Bi atoms to off-center in-plane. If we condense
the zone-boundary mode, we obtain the AFE $Pna2_1$ state. The AFE
mode, in turn, couples with the AFD$_{z}^+$ mode: even if the latter
is stable in this region of the BiFeO$_3$ phase diagram, it ``feels''
the large amplitude of the former, and is therefore brought slightly
off its equilibrium position. (In other words, we can regard the
AFD$_{z}^+$ distortion as a \emph{consequence} of the AFE mode, the
latter acting on the former as a ``geometric field''. The couplings
responsible for such an effect are described in
Section~\ref{couplings}.)
Conversely, if we go back to the reference structure and condense the
zone-center polar mode, there is no driving force whatsoever for an
AFD$_{z}^+$ distortion to appear. Instead, the uniform $\hat{P}_{xy}$
couples cooperatively with AFD$_{z}^-$. As before, the latter mode is
stable and would not occur by itself -- only as a secondary effect of
having $\hat{P}_{xy}$.  (The cooperative coupling is obvious from the
progressive increase of the AFD$_{z}^-$ amplitude for increasingly
large values of the in-plane electric displacement field,
$\hat{D}_{xy}$.)

\section{Discussion}

\subsection{Couplings}
\label{couplings}

The main couplings that are of relevance for this work are between
polarization and AFD modes, polarization and strain, and
antiferroelectricity and AFD modes. The ${\bf P}$-AFD coupling is well
known in BiFeO$_{3}$, as it tends to align the anti-phase AFD
pseudovector with the polar ${\bf P}$
vector.\cite{Kornev-06,Kornev-07}
In particular, consider the following term, which is quartic in the
mode amplitudes,
\begin{equation}
E^{{\rm AFD}-{\bf P}} = D \hat{P}_{xy} \hat{P}_{z} \omega^-_{xy} \omega^-_{z}.
\label{afdp}
\end{equation}
(Here $\omega_i$ indicates the amplitute of the $AFD^-_i$ mode,
following the notation of Refs.~\onlinecite{Kornev-06} and \onlinecite{Kornev-07}.
In our simulations along path~A we calculate, for $\hat{D}_{001} \sim
0$, large values of both $\hat{P}_{xy}$ and $\omega^-_{xy}$ (both tend
to a finite constant at $\hat{D}_{001}=0$). In such a regime,
Eq.~\eqref{afdp} states that $\omega^-_{z}$ must grow linearly with
$\hat{P}_{z}$ at small values of the latter, which is indeed what we
observe.
Along path~D, on the other hand, we start from a situation (at
$\hat{D}_{110} = 0$) where $\hat{P}_{z}$ and $\omega^-_{xy}$ are both
large, while the other two degrees of freedom vanish. Application of
an in-plane electric field induces a small $\hat{P}_{xy}$, which in
turn linearly induces a comparatively small $\omega^-_{z}$ due to the
coupling in Eq.~\eqref{afdp}.  (This coupling is precisely responsible for the coexistence of out-of-phase and in-phase octahedral rotations along the $z$ axis.)

Another important coupling of interest in the context of this work is
the well-known trilinear term in the AFE, $\omega^-_{xy}$ and
$\omega^+_{z}$,
\begin{equation}
E^{{\rm AFE}-\bm{\omega}} = C Q_{\rm AFE} \omega^-_{xy} \omega^+_{z}.
\label{afeomega}
\end{equation}
In presence of large $\omega^-_{xy}$, this term yields a cooperative
coupling between $Q_{\rm AFE}$ and $\omega^+_{z}$. This means that the
energy will be significantly lowered when the two instabilities
condense together (and thus possibly explain the surprisingly low
energy of the $Pnma$ phase\cite{Dieguez-11} -- it is only few meVs
higher than that of the $R3c$ ground state). Also, in those regions of
the phase diagram where only one of the latter two modes is an active
instability of the system, the stable mode will feel a ``geometric
field'' that forces it slightly off-center. We have seen examples of
such a behavior along both path~D and path~E: 
  in the former, a small
  $Q_{\rm AFE}$ amplitude persists even in a region of the phase diagram 
  where the $Q_{\rm AFE}$ mode is no longer ``soft'' 
  (i.e. for $\hat{D}_{110} > 1.25$), see Fig.~\ref{fig:pathd}(d); 
  in the latter, a small $\omega^+_{z}$ is present at $|\hat{D}_{110}| < 0.5$,
  see  Fig.~\ref{fig:pathe}(c)
  [the fact that $\omega^+_{z}$ is a secondary effect of $Q_{\rm AFE}$ 
  there can be inferred from the evolution of these two order parameters 
  along path~B].

\subsection{Behavior as a function of applied electric field}

\begin{figure*}
\begin{center}
\includegraphics[width=4.5in]{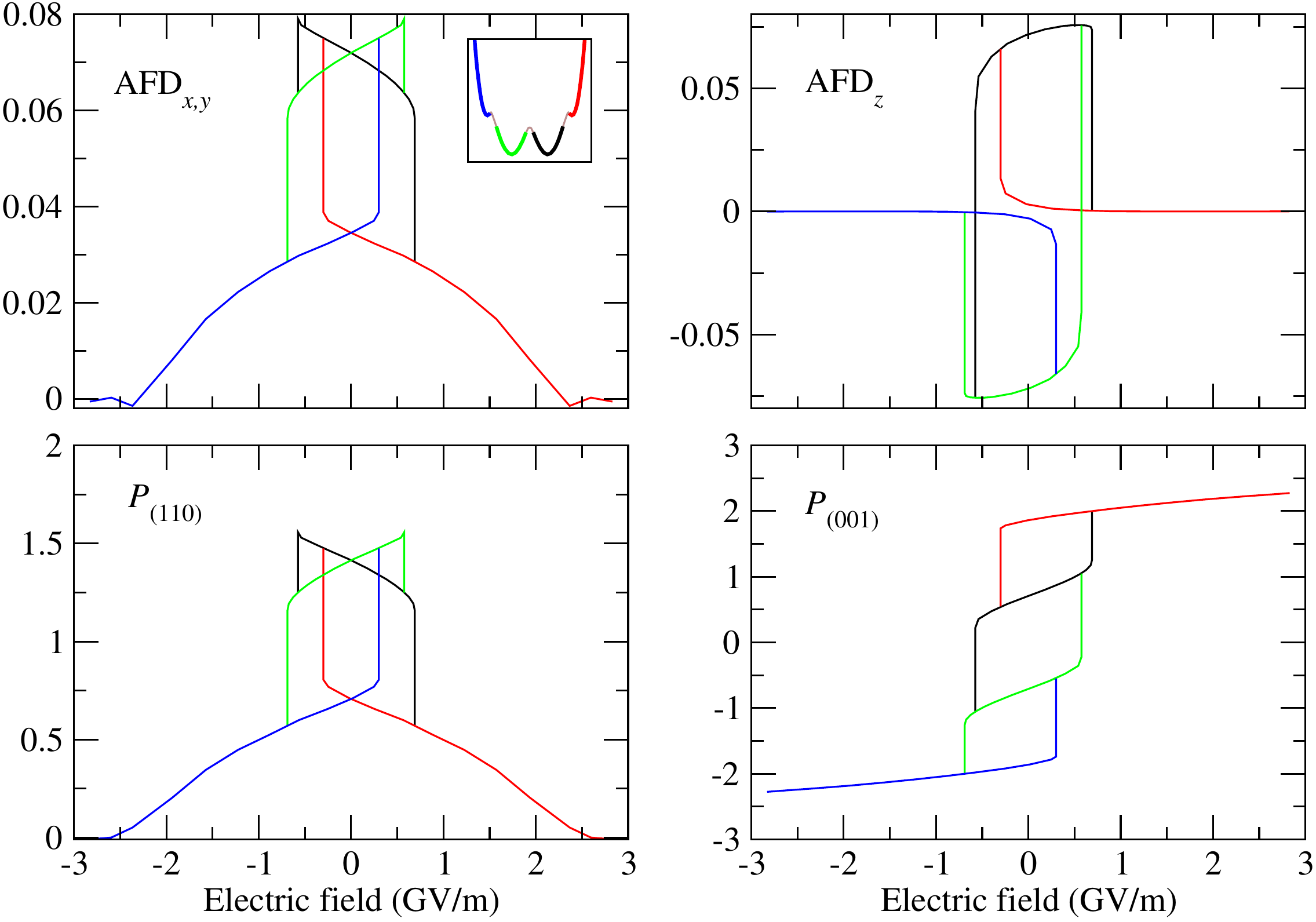}
\end{center}
\caption{ Switching via a [001]-oriented electric field. The four main
  panels represent the evolution of the relevant order parameters as a
  function of the applied field. The inset represents the potential
  energy curve [same as in Fig.~\ref{fig:patha}(a)], where the various
  segments have been highlighted according to the color code used in
  the main panels (unstable regions are marked in brown).
  Polarization and AFD amplitudes are in arbitrary units (consistent 
  with the fixed-$\hat{D}$ plots presented earlier).
\label{figswitch1}
}
\end{figure*}

We can use our results to inspect the behavior of BiFeO$_{3}$ as a
function of applied electric field, rather than the reduced electric
displacement.
This way we are able to simulate a hypothetical ferroelectric
switching experiment, performed at zero temperature and by
constraining the sample to remain in a monodomain configuration
throughout.
This also allows us to compare our results with those of previous
first-principles-derived approaches, e.g. those of
Ref.~\onlinecite{Lisenkov-09}.

We shall discuss two different switching paths, obtained respectively
by applying an electric field along the pseudocubic [001] or [110]
directions.
Interestingly, such information can also be used to access important
functional properties of BiFeO$_3$, and their evolution under an
external field of arbitrary magnitude.
For example, a recent experimental work has reported a significant
nonlinearity in the electromechanical response of BiFeO$_3$, with an
increase of the piezoelectric coefficient at high
fields;\cite{Chen-12} we shall address this topic in the last part of
this Section.

\subsubsection{Switching along [001]}

To represent the evolution of the system as a function of the electric
field, we start from the data discussed in Section~\ref{patha},
referring to path~A in electric displacement space.
First, we identify the segments where the system is in a (meta)stable
state, i.e. the parts of the curves where the electric field
\emph{increases} for increasing $\hat{D}_{001}$. (The remainder of the
data points are characterized by a negative capacitance, and as such
they correspond to unstable regions of the electrical phase diagram.)
Then, we replace the abscissas with the calculated value of the internal 
electric field, and finally we connect the different segments with vertical 
jumps wherever appropriate.
Note that such jumps always occur when a given phase reaches its limit
of (meta)stability, i.e. when a small increase in the electric field
magnitude produces a large variation in the structural parameters,
indicating a transition to a different phase.
These transitions are hysteretic, i.e. the system usually can be
switched back to the original phase, but at a different critical field
value.

\begin{figure*}
\begin{center}
\includegraphics[width=4.5in]{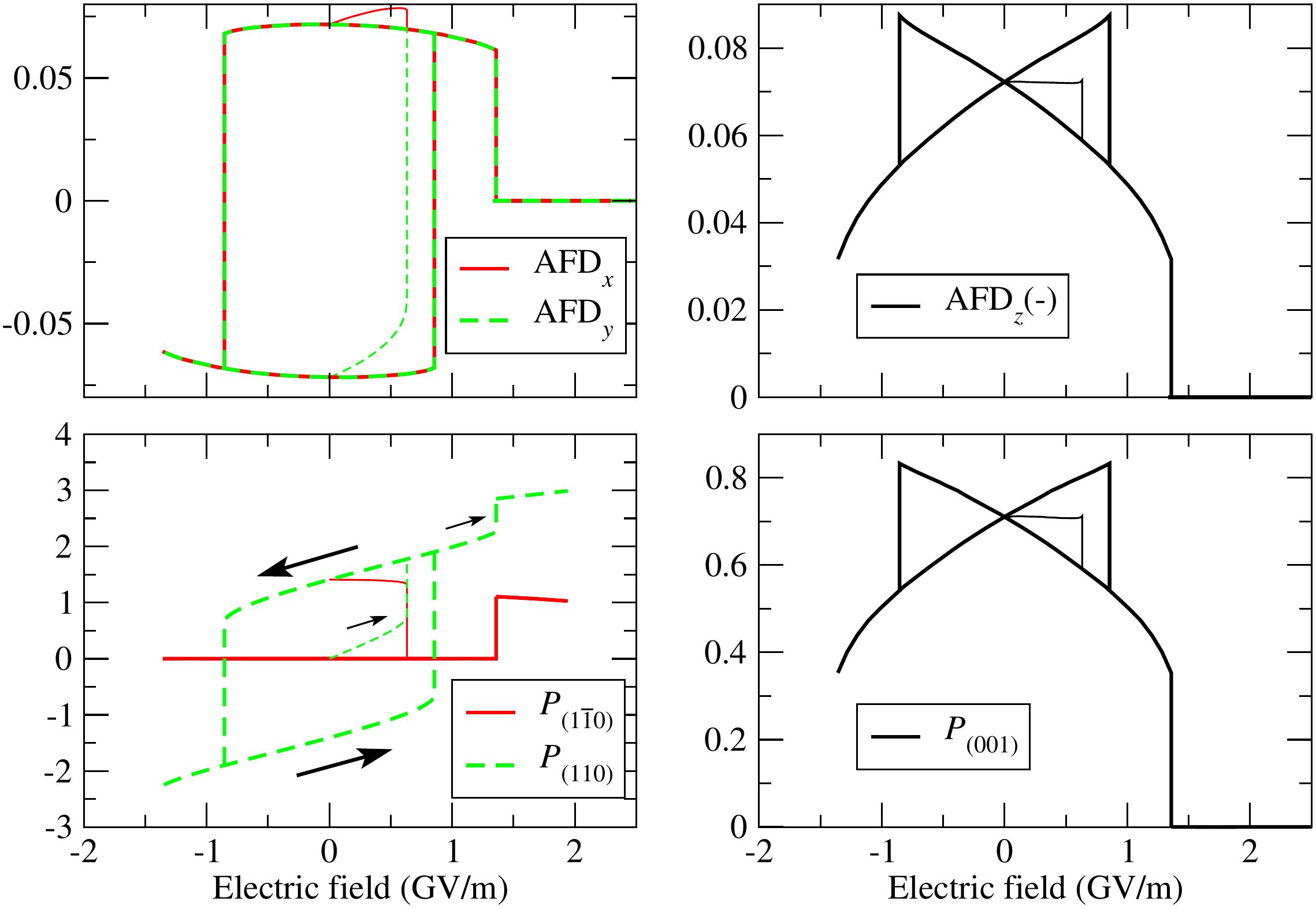}
\end{center}
\caption{ Switching via a [110]-oriented electric field. As the paths
  described by the system are uncomplicated to follow, we have
  reverted to the usual color code of the manuscript. Thinner lines
  refer to the evolution of the system when the initial state is the
  $\hat{D}_{110}=0$ configuration of Fig.~\ref{pathc}. Large arrows indicate
  the direction along which the system evolves within the main
  hysteresis loop. Small arrows indicate the evolution of the systems
  along secondary paths that lead into (or, alternatively, out from)
  the main loop. \label{figswitch2} 
  Polarization and AFD amplitudes are in arbitrary units (consistent 
  with the fixed-$\hat{D}$ plots presented earlier).
}
\end{figure*}

The resulting plots are shown in Fig.~\ref{figswitch1}.
In the four different panels we show, respectively, the AFD amplitude
(top) and the three components of the ``rigid-ion'' polarization; the
inset illustrates, as a guide, the color code that we used to
represent the different segments.
Note the multivalued character of the graphs, which is typical of
multistable ferroeoectric materials. At difference with common
ferroelectrics, however, BiFeO$_3$ displays four different states
at zero field, corresponding respectively to the two stable and
two metastable minima in the energy curve (inset).
As expected, the graphs referring to in-plane components are symmetric
with respect to an electric field inversion, while the out-of-plane
ones are antisymmetric.
At low to moderate fields, the AFD pseudovector behaves identically to
the polarization vector. Conversely, at high fields AFD$^-_z$ drops
abruptly to zero while the corresponding component of the polarization
keeps increasing ($Cc$-II phase).
At very high fields, also the in-plane AFD components go to zero, and
the system transitions to the $P4mm$ phase.
From our data, we can extract the four critical fields that will bring
the system through the following path: $Cc$-II$(-)$, $Cc$-I$(-)$,
$Cc$-I$(+)$, $Cc$-II$(+)$, $P4mm$$(+)$.
These are, respectively, 0.30~GV/m, 0.57~GV/m, 0.69~GV/m, and
$\sim$2.3~GV/m.
Note that the sequence of transitions is different from that of
Ref.~\onlinecite{Lisenkov-09}: the $Cc$-II phase was presumably not
known when Lisenkov \emph{et al.} constructed their effective
Hamiltonian, and the authors were therefore unaware of the two
corresponding metastable minima.
Even regarding the transition [$Cc$-I$(-)$ $\rightarrow$ $Cc$-I$(+)$]
that was studied in Ref.~\onlinecite{Lisenkov-09}, our results show
important differences at the quantitative level.
Most importantly, the critical field of Ref.~\onlinecite{Lisenkov-09}
appears to be largely overestimated with respect to ours ($\sim
1.5$~GV/m vs. our value of 0.57~GV/m).
This discrepancy probably originates from the approximations that are
typically used to build a simplified effective model. These consist in
discarding all the electronic and most of the lattice degrees of
freedom, preserving only few active variables in the problem.
As noted by others,\cite{Dieguez-11} we suspect that BiFeO$_3$ might
be a particularly difficult material in this context, because of the
strong nonlinearities induced by the lone-pair activity of Bi, which
lead to a remarkably rich and complex energy landscape.
Here many degrees of freedom interact in a highly nontrivial way,
making a simplified description particularly tricky.

It is interesting to note that the transition to a phase
with large aspect ratio ($Cc$-II) occurs, according to our results, at
a relatively small value of the electric field, 0.69~GV/m.
This may be within experimental reach (values up to 0.28~GV/m were
probed in Ref.~\onlinecite{Chen-12}), and thus open new opportunities
for the study of the supertetragonal phase of BiFeO$_3$.

\subsubsection{Switching along [110]}


Here we perform the same post-processing procedure described earlier,
only applying it to path~C [which refers to an electric field applied
  along the [110] direction] instead than path~A.
The results are shown in Fig.~\ref{figswitch2}.
When applying a [110]-oriented field to the $R3c$ structural ground
state, one can obtain different results depending on the orientation
of the spontaneous polarization in the initial state.
If we start with ${\bf P}\sim[1\bar{1}1]$, the field is initially
perpendicular to ${\bf P}$, and will tend to \emph{rotate} the
polarization vector towards the $[111]$ direction; the evolution of
the relevant structural degrees of freedom along this path is
illustrated by thin lines in Fig.~\ref{figswitch2}.
As we observed above, however, the rotation of ${\bf P}$ does not
proceed smoothly, but instead happens abruptly once the limit of
stability of the $[1\bar{1}1]$-oriented phase is attained.
In particular, at first $P_{001}$, $P_{1\bar{1}0}$ and AFD$^-_z$
remain constant while both $P_{110}$ and AFD$^-_{x,y}$ increase
linearly. For larger fields some nonlinearities show up, until
eventually, at a critical field of 0.63~GV/m, the system transitions
to the [111]-oriented structure.
[Note that, once such a transition has occurred, the system will not
  go back to the $[1\bar{1}0]$-oriented state, hence the absence of a
  corresponding hystheresis loop in the plots.]
For larger fields, the polarization component that is collinear with
the field, $P_{110}$, keeps increasing at the expense of $P_{001}$
and AFD$^-_z$; in the same interval AFD$^-_{x,y}$ remain roughly
constant (actually, our results show a slight decrease).
At a second critical field of 1.36~GV/m, the system switches to the
$Pm$ state described in Section~\ref{pathc}, where the AFD distortions
disappear completely and a large polarization develops along an
off-axis in-plane direction.

If after the first transition, from the $[1\bar{1}1]$- to the
[111]-oriented phase, instead of further increasing the electric field
we decrease it again to zero, we end up in the [111]-oriented $R3c$
ground state.
From here, by applying a negative field, we obtain the opposite result
compared to above, i.e. $P_{110}$ \emph{decreases} while both
$P_{001}$ and AFD$^-_z$ simultaneously increase.
At a sufficiently large negative field of $-$0.85~GV/m, the system
switches directly to a $[\bar{1} \bar{1} 1]$-oriented state; as we
said above, this happens without visiting ever again the $[1\bar{1}1]$
region of the phase diagram.
Here, AFD$^-_{x,y}$ and $P_{110}$ both switch sign, while
$P_{001}$ and AFD$^-_z$ remain positive, in spite of undergoing a
significant reduction.

This latter observation is in qualitative disagreement with the
simulations of Ref.~\onlinecite{Lisenkov-09}, where \emph{all} the
components of ${\bf P}$ were found to switch simultaneously when a
sufficiently large [110]-oriented field is applied to a $[\bar{1}
  \bar{1} 1]$-type $R3c$ starting point.
The reasons behind such a discrepancy are presently unknown. We can
only speculate that the dynamics of the transition might play a role
(in our calculations we evolve the system quasistatically along the
switching path, while some thermal activation is considered in
Ref.~\onlinecite{Lisenkov-09}). Or, alternatively, subtle differences in
the potential landscape (possibly associated with the effective
Hamiltonian generation procedure) might be the real culprit. Further
calculations and cross-checks will be necessary to clarify this point.
Apart from this, the critical fields that we obtain here appear to be,
again, much smaller than those calculated by Lisenkov {\em et al.} For
example, reversal of the in-plane polarization occurs in our
calculations at a field of 0.85~GV/m, compared to the value of
1.62~GV/m reported in Ref.~\onlinecite{Lisenkov-09}.
At higher fields, Lisenkov {\em et al.} report a transition to a
$Ima2$ state, which in our calculations never occurs, at a field of
4.26~GV/m, and a second transition to $Pm$ at 5.5~GV/m.
We find, instead, that the system transitions directly to $Pm$,
without passing through the $Ima2$ state (the $Ima2$ phase appears to
be unstable at any value of the electrical displacement field in our
calculations), and at a substantially smaller field of 1.36~GV/m.
Such a large difference in the critical fields suggests that the
potential landscapes of our BiFeO$_3$ model and that of
Ref.~\onlinecite{Lisenkov-09} significantly differ, which would justify
the aforementioned discrepancy in our respective switching paths.

\subsubsection{Piezoelectric coefficient at high field}

\begin{figure}
\begin{center}
\includegraphics[width=3.2in]{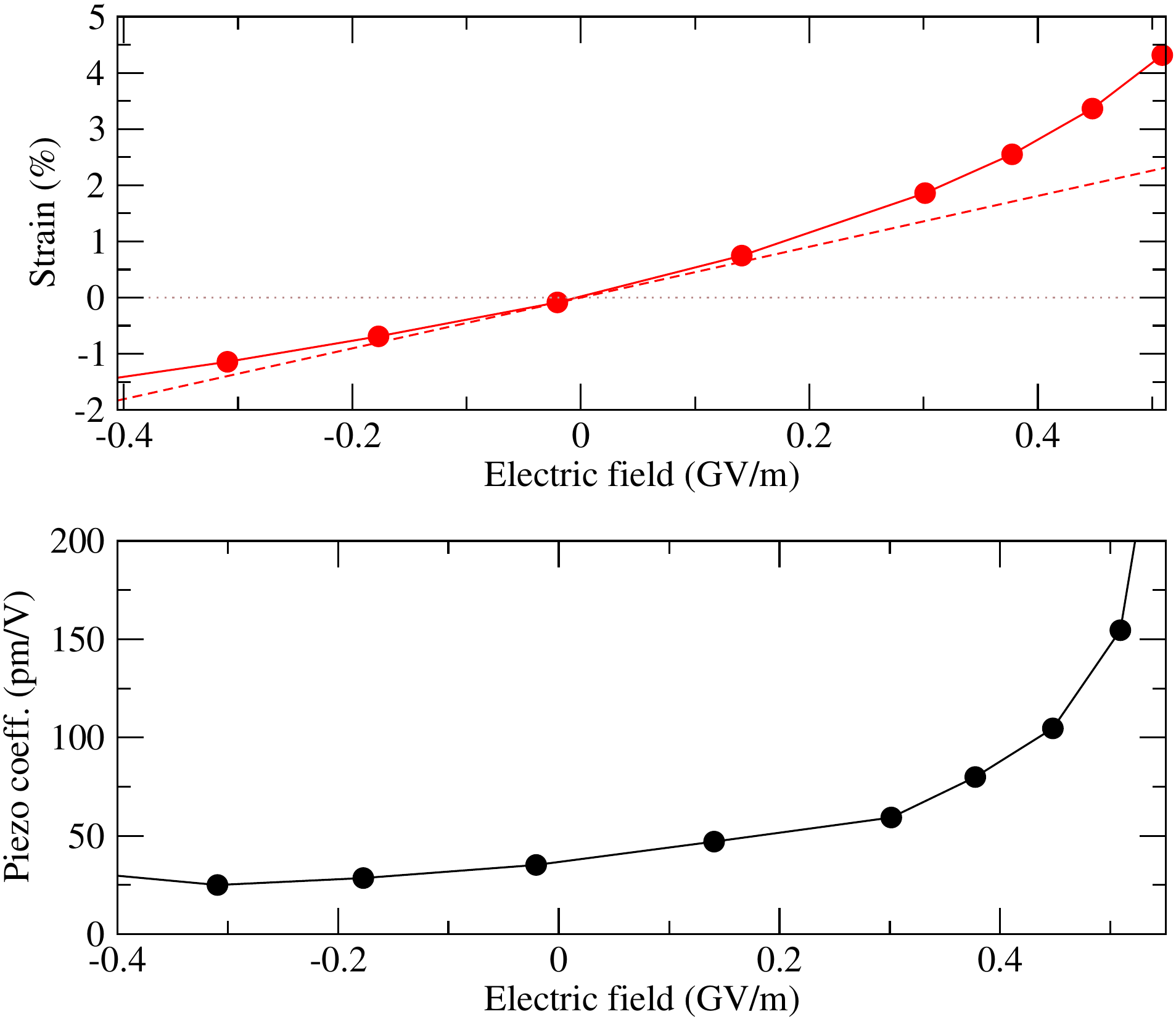}
\end{center}
\caption{ Piezoelectric coefficient as a function of the applied
  electric field. \label{figpiezo} }
\end{figure}

Chen {\em et al.}\cite{Chen-12} have recently studied the
piezoelectricity of a BiFeO$_3$ thin film, and found that at high
electric fields the response deviates significantly from the low-field
values.
In particular, the out-of-plane strain component displays a strong
nonlinearity, reaching values of up to 2\% at the highest fields
(0.28~GV/m); consequently, the piezoelectric constant increases from
55~pm/V up to 86~pm/V.
This outcome was ascribed to the lattice softening that occurs in the
proximity of a phase transition; the electric field would drive the
system close to the $Cc$-II state, and thus induce the observed
nonlinearities.

Verifying such a scenario is relatively straightforward with the data
that we already have in our hands.
It suffices, along path~A, to express the $c$ parameter of the cell as
a function of the applied potential, $V$; then, the piezoelectric
coefficient is readily given by differentiating the former with
respect to the latter. (Strictly speaking, our data should
be compared with some caution to those of Ref.~\onlinecite{Chen-12}:
in the latter the in-plane lattice parameters are clamped to the
substrate, and do not show any variation with the applied field; in
our computational experiment, all structural parameters of the cell
are left free to relax.)

The results for the equilibrium out-of-plane strain and piezoelectric
constant are shown, as a function of the applied field in
Fig.~\ref{figpiezo}.
At high values of the field, the system undergoes a drastic increase
of the $c$ parameter, reaching a strain of 4\% at $E \sim$0.5~GV/m; at
0.3~GV/m the strain is about 2\%, in nice agreement with the
experimental data of Ref.~\onlinecite{Chen-12}.
Such a strongly nonlinear behavior (the hypothetical linear regime is
represented by a thin dashed line, for comparison) is clearly a
consequence of lattice softening in proximity of the $Cc$-II phase,
confirming the arguments of Chen {\em et al.} -- as we said, the
transition itself occurs in our calculations at fields (0.69~GV/m)
that are only slightly larger than those shown in Fig.~\ref{figpiezo}.
The piezoelectric coefficient at zero field is significantly smaller
in our calculations than in Ref.~\onlinecite{Chen-12}: 37~pm/V versus
55~pm/V. (This may be a consequence of our use of the LDA
approximation, which tends to harden the phonon frequencies and hence
the lattice response to a field.) Nevertheless, the field-induced
increase appears to be in excellent agreement: from 37~pm/V to 60~pm/V
in our calculations, versus 55~pm/V to 86~pm/V in the experiment,
i.e., about 60\% in both cases.

\subsection{Stability under an applied field}

\begin{figure}[b!]
\begin{center}
\includegraphics[width=3.2in]{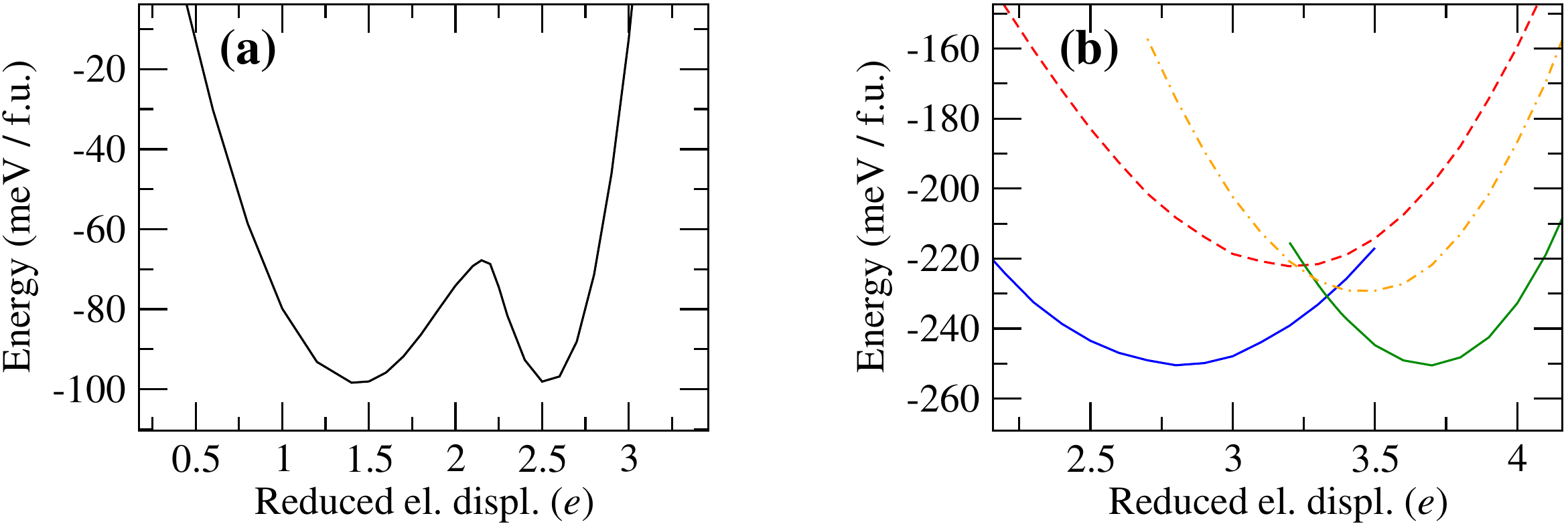}
\end{center}
\caption{ Phase stability as a function of the applied electric field.
  (a): energy versus a [001]-oriented electric displacement under an
  applied external voltage $V_{\rm ext}$~=~0.33~V. (b): energy versus
  a [110]-oriented electric displacement under an applied external
  voltage $V_{\rm ext}$~=~0.41~V.
\label{figvext}
}
\end{figure}

In the examples above, the transition between different phases was
determined by the stability range of the starting point, without
regards for the relative energetics of the final configuration.
Because of the assumption of an ideal monodomain crystal, the
resulting critical fields are largely overestimated, e.g. with respect
to the experimental coercive fields.
Defects and/or inhomogeneities in a real material typically facilitate
switching by providing nucleation centers and alternative transition
paths (e.g. via domain wall motion).
This needs to be kept in mind when comparing our results with the
available experimental data.

From our data there, however, one can extract other pieces of
information that are, in principle, much less sensitive to the above
issue.
In particular, one may be interested in the \emph{stability range} of
a given phase, rather than in the actual transition path that leads
from one phase to another.
More concretely, one can ask the following question: ``What voltage do
I need to apply to the sample in order for a given phase to be the
structural ground state?''
In the context of our fixed-$D$ calculations, answering this question
is particularly easy -- it only involves operating a simple
modification to the total energy curves that takes into account the
applied external voltage.
This implies, in practice, adding a linear function to the internal
energy,
\begin{equation}
U'(\hat{D}) = U(\hat{D}) + V_{\rm ext} \hat{D},
\end{equation}
where $V_{\rm ext}$ is the external bias applied to the relevant facet
of the simulation cell. The critical voltage corresponds then to the
value of $V_{\rm ext}$ for which two configurations, belonging to two
separate segments of the electrical equation of state, become
degenerate in energy.
(The situation is in all respects analogous to the study of phase
stability under applied mechanical stress; the electric displacement
field can be thought as the ``volume'' coordinate, whereas the applied
voltage is the counterpart of the hydrostatic pressure.)

In Fig.~\ref{figvext} we show the results of such an analysis, which
we performed in the two cases described in the previous Section of
[001]-oriented and [110]-oriented voltage drops, respectively.
An external voltage of 0.33~V/cell is sufficient to make the
supertetragonal $Cc$-II phase degenerate with the lowest-energy point
in the $Cc$-I region.
This is, therefore, the minimum voltage that is necessary for the
high-field $Cc$-II phase to become accessible; it corresponds to an
electric field of about 400~MV/m and is, therefore, very close to the
experimentally accessible range.
The corresponding analysis of the [110]-oriented case leads to a
stability limit of the $Pm$ phase of 0.41~V/cell, corresponding to a
comparatively larger electric field of 700~MV/m. Note that this
critical value is, in any case, significantly smaller than 1.36~GV/m,
which we found in the previous Section when studying the actual
transition.

\section{Conclusions}

We have studied the electrical phase diagram of bulk BiFeO$_3$ by
describing four different paths in electric displacement space.
Four important aspects of the physics of BiFeO$_3$ emerge from the
results presented here: (i) the presence of nontrivial couplings
between the polar and other structural degrees of freedom; (ii) the
possibility of controlling antiferrodistortive modes of BiFeO$_3$ via
application of an electric field; (iii) the relevance of
antiferroelectricity in some regions of the electrical phase diagram,
where it leads to a characteristic triple-well potential as a function
of $\hat{D}_{110}$ [one can regard (ii) and (iii) as consequences of
  (i)]; (iv) the five-well potential of BiFeO$_3$ as a function of
$\hat{D}_{001}$, which results from the presence of a metastable
$Pnma$ phase at small values of $\hat{D}_{001}$, and of a high-aspect
ratio $Cc$ phase at large values of $\hat{D}_{001}$.

The above results provide original physical insight into the complex
energy landscape of BiFeO$_3$, which can be taken as a model system to
study the interplay between different structural modes and, in
particular, the competition between ferroelectric and
antiferroelectric phases. Our results also constitute a stringent
benchmark for future development of approximate (e.g. effective
Hamiltonian) models.

\acknowledgements

Work funded by MINECO-Spain Grant Nos. FIS2013-48668-C2-2-P (MS) and MAT2013-40581-P
(JI), by Generalitat de Catalunya Grant 2014 SGR 301 (MS), and by 
FNR Luxembourg Grant FNR/P12/4853155/Kreisel (JI).

\bibliography{bifeo3}

\end {document}